\documentclass[iop]{emulateapj}

\usepackage{epsfig}
\usepackage{amsmath}
\usepackage{adjustbox}

\newcommand{\lyansp}{Ly-$\alpha$}

\newcommand{\cii}{C~{\sc ii}~}

\newcommand{\civ}{C~{\sc iv}~}

\newcommand{\oii}{[O~{\sc ii}]~}

\newcommand{\siii}{Si~{\sc ii}~}

\newcommand{\hi}{H~{\sc i}~}

\newcommand{\mgiinsp}{Mg~{\sc ii}}

\newcommand{\ciinsp}{C~{\sc ii}}
\newcommand{\oinsp}{O~{\sc i}}
\newcommand{\siiinsp}{Si~{\sc ii}}

\newcommand{\mginsp}{Mg~{\sc i}}
\newcommand{\mgii}{Mg~{\sc ii}~}

\newcommand{\angmath}{\textrm{\AA}}

\newcommand{\mmgii}{{\rm Mg \; \mbox{\tiny II}}}

\newcommand{\kms}{km s$^{-1}$}

\newcommand{\NumSystems}{280}
\newcommand{\NumQSOs}{100}
\newcommand{\NumExtras}{54}
\newcommand{\NumHighZ}{7}

\begin{document}


\title{\ion{Mg}{2} Absorption at $2 < z < 7$ with Magellan/FIRE, III. Full
  Statistics of Absorption towards \NumQSOs ~High-Redshift QSOs\altaffilmark{1}}

\author{Shi-Fan S. Chen\altaffilmark{2,7,*}, 
Robert A. Simcoe\altaffilmark{2,6}, 
Paul Torrey\altaffilmark{2,8}, 
Eduardo Ba\~nados\altaffilmark{3}, 
Kathy Cooksey\altaffilmark{4}, 
Tom Cooper\altaffilmark{2},
Gabor Furesz\altaffilmark{2},
Michael Matejek\altaffilmark{2},
Daniel Miller\altaffilmark{2},
Monica Turner\altaffilmark{2},
Bram Venemans\altaffilmark{5},
Roberto Decarli\altaffilmark{5},
Emanuele P. Farina\altaffilmark{5},
Chiara Mazzucchelli\altaffilmark{5},
Fabian Walter\altaffilmark{5}}
\altaffiltext{1}{This paper includes data gathered with the 6.5 meter
  Magellan Telescopes located at Las Campanas Observatory, Chile.}
  \altaffiltext{2}{MIT-Kavli Center for Astrophysics and Space
    Research, Massachusetts Institute of Technology, 77 Massachusetts
    Ave., Cambridge, MA 02139, USA}
  \altaffiltext{3}{Observatories of the Carnegie Institution for Science}
  \altaffiltext{4}{University of Hawai`i at Hilo}
  \altaffiltext{5}{Max Planck Institute for Astronomie}
  \altaffiltext{6}{Radcliffe Institute for Advanced Study}
  \altaffiltext{7}{Department of Physics, University of California, Berkeley, CA 94720, USA }
  \altaffiltext{8}{Hubble fellow}
  \altaffiltext{*}{shifan\_chen@berkeley.edu}

\begin{abstract}

We present statistics from a survey of intervening \mgii
absorption towards \NumQSOs ~quasars with emission redshifts between
$z=3.55$ and $z=7.09$.  Using infrared spectra from Magellan/FIRE, we
detect \NumSystems ~cosmological \mgii absorbers, and confirm that the
comoving line density of $W_r>0.3$ \AA ~\mgii absorbers does not evolve measurably between $z=0.25$ and $z=7$.  This is
consistent with our detection of seven \mgii systems at $z>6$, redshifts not covered in prior searches.  Restricting to systems with $W_r>1$\AA, there is significant evidence for
redshift evolution.  These systems roughly double in density
between $z=0$ and $z=2$-$3$, but decline by an order of magnitude from
this peak by $z\sim 6$.  This evolution mirrors that of the global
star formation rate density, potentially reflecting a connection between
star formation feedback and the strong \mgii absorbers.
We compared our results to the Illustris cosmological simulation at z=2-4 by assigning absorption to catalogued dark-matter halos and by direct extraction of spectra from the simulation volume. Reproducing our results using the former requires circumgalactic MgII envelopes within halos of progressively smaller mass at earlier times. This occurs naturally if we define the lower integration cutoff using SFR rather than mass. Spectra calculated directly from
Illustris yield too few strong \mgii absorbers.  This may arise from unresolved phase space structure
of circumgalactic gas, particularly from spatially unresolved turbulent or bulk motions.  The presence of circumgalactic magnesium at $z>6$ suggests that enrichment of intra-halo gas may have begun
before the presumed host galaxies' stellar populations were mature and
dynamically relaxed.
\end{abstract}

\section{Introduction}

For over 30 years \citep{bergeron86,bergeron_boisse,bahcall_spitzer},
the \mgii doublet has been recognized as an absorption signature of
enriched gas in the halos of luminous galaxies.  While most Mg is
singly ionized in the Galactic disk on account of the 0.56 Ryd
ionization energy of \mginsp, blind absorption surveys predominantly
identify discrete \mgii absorbers (e.g. above an equivalent width
threshold of $W_r\sim 0.1-0.3$\AA) in the more extended halos of
distant galaxies at impact parameters of 10-100 kpc, or a few tenths
of $R_{vir}$ \citep{chen2010lowz,
  zibetti2007colors,bouche2007ha,gauthier2010,lovegrove2011,
  churchill_2000a,
  werk_COS_emp, churchill_rvir}.  Gas at these impact parameters presents a larger
cross section for chance absorption, yet retains pockets of sufficient
density that \hi can shield \mgii ions against photons with their
ionization energy of 1.1 Ryd.

The empirical association of galaxies with intra-halo \mgii gas,
together with the heavy element enrichment implied by Mg, invites the
interpretation that \mgii absorption arises in regions polluted by
galactic winds.  This is an attractive picture because simulations of
galaxy formation require vigorous amounts of mechanical and thermal
feedback to match galaxies' stellar mass function and mass-metallicity
relation \citep{illustris}, and the halo is a convenient place to
deposit baryons ejected from the disk during this process.  
Unfortunately these same simulations
are not always well-suited to make detailed predictions of \mgii
properties of circum-galactic gas.  In regions of the
temperature-density plane where the \mgii ionization fraction peaks,
numerical codes often transition into sub-grid scalings for cooling
and mass flow \citep{illustris2}.  

Simple analytic calculations of the total circum-galactic mass and 
metal budget from observations of projected galaxy-QSO or QSO-QSO 
pairs derive very large masses
\citep{tumlinson_reservoir, bordoloi_carbon, stern_universal,
  QPQ_mass}, despite the fact that ionization models for individual
optically thick absorbers consistently yield line-of-sight sizes
measured in tens of physical parsecs
\citep{charltonetal03,misawaetal07,lynchcharlton07,simcoe06_cgm,
stern_universal}.
This is corroborated by observations of \mgii absorption in lensed
QSOs which show variations
in low ionization absorption (\mgiinsp, \siiinsp, and \ciinsp) on transverse 
scales ranging from 26$h^{-1}$ pc \citep{rauch99} to 
200-300 $h^{-1}$ pc \citep{rauch_lens}.  These findings suggest that \mgii
absorbing gas is highly structured in halos even as observations of
the high covering fraction show that it is widespread.

Further complicating the picture from simulations, the halo is
expected to harbor accreting gas at similar densities, both on first
infall from the IGM
\citep{dekel,keres,faucher_coldflows,fumagalli_flows}, and recycled
from previous generations of star forming winds that remained bound to
the dark matter halo \citep{oppenheimer,ford_cos}.

Indeed, infalling \mgii absorption has been seen directly in down-the-barrel
spectra of selected nearby galaxies \citep{rubin2012}, in contrast to
the common outflowing/blueshifted \mgii seen in stacks of galaxy
spectra at similar redshift \citep{ubiquitous2009}.  Apparently galaxy
halos contain \mgii gas from both inflowing and outflowing baryons in
unknown proportion.  Morphological analysis of absorber host galaxies
lends tentative evidence to this hypothesis, since strong absorption
is slightly more likely out of the disk plane, while weaker absorbers
can align with the orientation of the disk
\citep{bouche2007ha,bordoloi,kacprzak2011incl,nielsen_orientation}.

Models of accretion flows and galactic winds both exhibit redshift
dependence, but \mgii observations in optical spectrographs probe a
maximum absoprtion redshift of $z\sim 2.5$. In \citet[][hereafter
  Paper I]{PaperI}, we presented initial results on an infrared survey
for \mgii absorbers at $2<z<5.5$, using the FIRE spectrograph on
Magellan \citep{fire_pasp}.  Out of necessity, the
IR sample is much smaller than optical \mgii surveys, which include up
to $\ge 30,000$ doublets
\citep{nestor2005,prochtersdss2006,lundgren2009,quiderSDSS,seyffert,zhu_mgii,chen_boss_mgii,dr12_mgii}.
Since \mgii appears to trace both star-formation feedback
\citep{bond,ubiquitous2009,menardo2,zibetti2007colors,bouche2007ha,nestor2010strong,
  crystal_outflows,kornei} and cool accretion
\citep{mgii_rotation,keres,rubin2010,lopezChen,muse_coldflow}, our aim
was to extend redshift coverage past the peak in the star formation
rate density, providing statistics on absorption during the buildup
phase of stellar mass. 

For robust statistics, our goal was to observe $\sim 100$ QSOs with
FIRE and identify 100-200 absorbers.  Paper I presented the first 46
sightlines, limited by observing time and weather.  Here we update
these results to include \NumExtras ~additional sightlines for a total
sample of \NumQSOs ~objects, constituting the full survey.

Paper I focused on bright QSOs to build up the sample; a consequence
of this choice is that our statistics were best at $2<z<4$ because of
the abundance of bright background sources.  A key result of this
early paper was evidence for evolution in the frequency of strong
\mgii absorbers ($W_r>1.0$\AA), which peak in number density near
$z\sim 2.5$ and then decline toward higher redshift.  The significance
of this result hinged on decreasing numbers of strong \mgii in the
highest redshift bins, which contained less survey pathlength because
the highest redshift ($z>6$) background sources are rarer and fainter.

In the intervening time, new wide-area surveys with near-IR color
information have yielded numerous examples of bright $z>5.5$ QSOs in
the Southern Hemisphere and therefore accessible for FIRE observation
\citep{banados_panstarrs,
  jiang_sdss_final,venemans_viking,venemans_panstarrs,willott_cfhqs,venemans13,banados14}.
These sightlines are suitable for \mgii absorption surveys and this
paper employs a larger proportion of observing time on them, with a goal of improving statstics at $z>4$. By emphasizing these high-redshift targets we also obtain greater overlap with pioneering investigations cool absorbing gas at $z>6$ \citep{becker_oi,becker_2011} that focused on low-ionization \oinsp, \ciinsp, and \siii visible in high-resolution optical spectra. These authors speculate that cool absorbing gas populates the circum-galactic media of galaxies that are too faint to observe at present, but which are thought to be important for hydrogen reionization.

We employ largely the same analysis techniques as Paper I utilizing
the new and larger \mgii sample.  In Sections 2 and 3 we describe the methods for data collection,
continuum fitting, line finding, and tests for completness and sample
contamination from false-positives.  Section 4 presents updated
results on the line density and evolution of \mgii frequency and
absorber equivalent width distributions.  Section 5 discusses these
results in the context of different models for \mgii production.  For
comoving calculations we assume a cosmology derived from
the the Planck 2016 results with $H_0=67.27,
\Omega_M=0.3156, \Omega_\Lambda=1 - \Omega_\Lambda$ throughout \citep{planck2015}.

\section{Data}

This paper expands the original \mgii survey of Paper I to
\NumQSOs ~sightlines, adding \NumExtras ~objects to our original
sample of 46.  This achieves our original goal of surveying $\sim 100$
QSOs, while focusing more heavily on quasars with high emission
redshift.  This approach carries a larger observational cost, but was
motivated by the findings of Paper I---specifically, that the
strongest \mgii absorbers decline in frequency above $z\sim 3$ but
weaker systems with $W_r=0.3-1.0$\AA ~remain nearly constant in
comoving number density.  These results hinged on the highest redshift
bins of the original sample, which had the shortest absorption path
and therefore the highest uncertainty.

Our sightlines are drawn from a number of quasar surveys.  The
majority of the sample is drawn from the SDSS DR7 QSO catalog
\citep{schneider} and dedicated high redshift SDSS searches
\citep{jiang_sdss_final}, but significant numbers are also derived
from the BR and BRI catalogs which contain many Southern APM-selected
quasars \citep{sl1996}.  Many of the new $z>6$ sightlines observed for
this paper are drawn from searches for $i$ and $z$ dropouts in the
UKIDSS, PanStarrs, and VISTA/VIKING surveys \citep{mortlock,
  venemans_panstarrs, venemans_viking,
  banados_panstarrs,willott_cfhqs,venemans13,banados14,venemans_inprep,mazzucchelli_inprep},
which now have discovered a significant fraction of all known $z>5.5$
QSOs.  Objects were selected for observation based on the QSO's
redshift and apparent magnitude.  No consideration was given to the
intrinsic properties of the background objects other than a screening
to avoid broad absorption line (BAL) quasars, which contain extended
intrinsic absorption that can be confused with intervening,
cosmological lines.

All obervations were conducted with FIRE, which is a single object,
prism cross-dispersed infrared spectrometer on the Magellan Baade
telescope \citep{fire_pasp}.  We observed with a 0.6$^{\prime\prime}$
slit, yielding a spectral resolution of $R=6000$, or approximately 50
\kms, over the range 0.8 to 2.5 $\mu$m.  A complete list of these QSOs
may be found in Table \ref{tab:qsoList}.  The spectra were reduced
using the IDL {\tt FIREHOSE} pipeline, which performs 2D sky
subtraction using the algorthms outlined in \citet{kelson_optimal} and
extracts an optimally weighted 1D spectrum.  Telluric corrections and
flux calibration are performed using concurrently observerd A0V
standard stars, which are input to the {\tt xtellcor} routine drawn
from the {\tt spextool} software
library\citep{cushing2004spextool,vacca2003tell}.  The signal-to-noise
ratios (SNRs) of the spectra vary substantially and are indicated in
Table 1; these differences are accounted during the completeness
corrections outlined in Section 3.

\begin{figure}[t]
	\begin{minipage}{0.5\textwidth}
	\includegraphics[width=\textwidth]{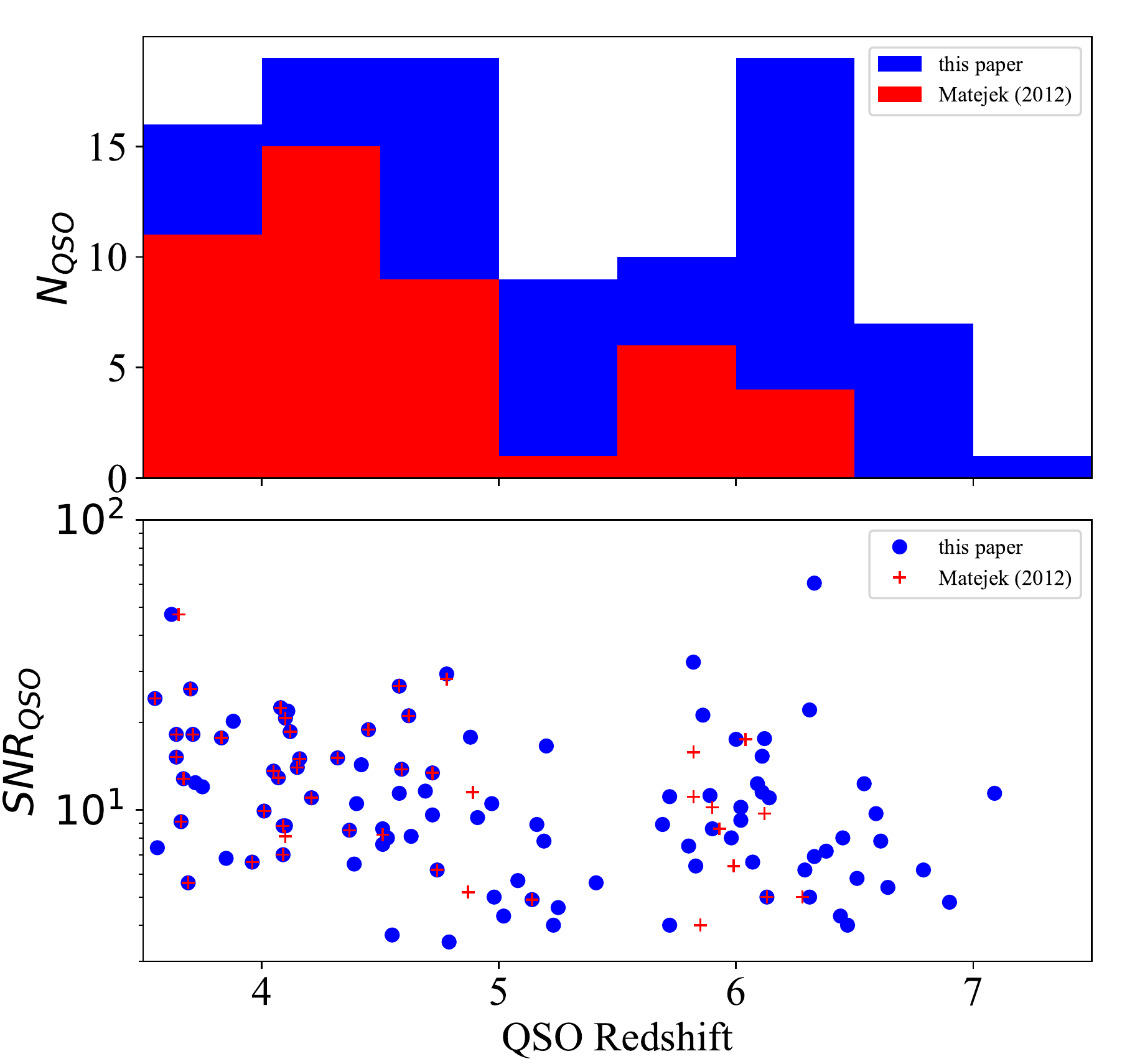}
	\end{minipage}
\caption{(Top) A histogram of our quasar sample by emission redshift
          (blue) as compared to the original sample (red), in bins
          of $\Delta z = 0.5$. The number of QSO's studied beyond z =
          5.5 has been doubled, and the median emission redshift has
          been increased from z = 4.27 to z = 4.63. (Bottom) A scatterplot showing the emission redshift and signal-to-noise ratios of our quasar spectra as compared to \citet{PaperI}.}
\label{fig:qsohist}
\end{figure}

The lower redshift limit for our \mgii absorption search is $z \approx
1.9$, set by the wavelength coverage of FIRE as in the original
survey. The upper limit, fixed at 3000 km s$^{-1}$ below the emission
redshifts of the QSOs, has been significatly increased. Whereas the
maximum QSO emission redshift in original survey was $z=6.28$ (set by
SDSS1030+0524), our sample now includes six QSOs with emission redshift
$z > 6.5$, including ULAS1120+0641 at $z=7.09$
(Fig.~\ref{fig:qsohist}).

Despite having several objects with emisison redshifts $z_{em}>6$, the
original survey had an {\em absorption} redshift limit of only
$z_{abs}=5.4$ even though we observed several quasars at $z\gtrsim 6$.
This reflects limitations from atmospheric absorption between the $H$
and $K$ bands, which cuts out \mgii pathlength from $5.4<z<5.9$.
Paper I included too few objects with coverage above $z=6$ to derive
meaningful constraints on \mgii in this epoch. For this paper, we have
therefore dedicated the majority of our observing time to fainter QSOs
at $z = 6$ and above, thereby increasing our constraining power on the
column density of \mgii absorbers at these at earlier redshifts, rather
than strictly maximizing the total number of QSOs observed. Our
expanded sample has more than doubled the pathlength above $z=5.5$,
and the median emission redshift has increased from $\langle z\rangle=4.27$ to
$\langle z\rangle=4.63$.

\begin{deluxetable*}{l c c c c c c c}[!htbp]
\tablecaption{FIRE \mgii Survey Sightlines}
\tablehead{ \colhead{Quasar}  & \colhead{$z_{em}$} & \colhead{$\Delta z$} & \colhead{$t_{exp}$ } & \colhead{Median SNR\tablenotemark{a}} & \colhead{RA} & \colhead{DEC} \\ \colhead{} & \colhead{} & \colhead{} & \colhead{(sec.)} & \colhead{(pixel$^{-1}$)}  }
\startdata
Q0000-26 & 4.10 & 1.95-3.83 & 1226 & 20.7 & 00:03:22.9  & -26:03:18.8
 \\

BR0004-6224 & 4.51 & 1.95-4.51 & 1764 & 7.6 & 00:06:51.6  &  -62:08:03.7
 \\

BR0016-3544 & 4.15 & 1.95-3.83 & 2409 & 14.0 & 00:18:37.9  &  -35:27:40.
 \\

SDSS J0040-0915 & 4.97 & 1.95-4.97 & 2409 & 10.5 & 00:40:54.65  &  -09:15:26.0
 \\

SDSS J0042-1020 & 3.88 & 1.95-3.83 & 4818 & 20.2 & 00:42:19.74  &  -10:20:09.5
 \\

SDSS J0054-0109 & 5.08 & 1.95-5.08 & 4501 & 5.7 & 00:54:21.43  & -01:09:21.0
 \\

SDSS J0100+2802 & 6.33 & 2.19-6.33 & 18652 & 60.5 & 01:00:13.02 & +28:02:25.84
 \\

SDSS 0106+0048 & 4.45 & 1.95-4.45 & 3635 & 18.9 & 01:06:19.2  &  +00:48:23.
 \\

VIK J0109-3047 & 6.79 & 2.39-6.79 & 28511 & 6.2 & 01:09:53.1  &  -30:47:26.3
 \\

SDSS J0113-0935 & 3.67 & 1.95-3.67 & 1944 & 12.8 & 01:13:51.96  &  -09:35:51.1
 \\

SDSS J0127-0045 & 4.08 & 1.95-3.83 & 3635 & 22.5 & 01:27:00.69  &  -00:45:59.2
 \\

SDSS J0140-0839 & 3.71 & 1.95-3.71 & 1226 & 18.2 & 01:40:49.18  &  -08:39:42.5
 \\

SDSS J0157-0106 & 3.56 & 1.95-3.56 & 1817 & 7.4 & 01:57:41.57  & -01:06:29.6
 \\

PSO J029-29     & 5.99 & 2.04-5.98 & 4501 & 8.0 & 01:58:04.14  & -29:05:19.25
 \\

ULAS J0203+0012 & 5.72 & 1.95-5.40 & 3635 & 4.0 & 02:03:32.38  & +00:12:29.27
 \\

SDSS J0216-0921 & 3.72 & 1.95-3.72 & 1920 & 12.4 & 02:16:46.9 & -09:21:07.0
 \\

SDSS J0231-0728 & 5.41 & 1.95-5.40 & 2409 & 5.6 & 02:31:37.6  & -07:28:54.0
 \\

SDSS J0244-0816 & 4.07 & 1.95-3.83 & 1944 & 12.9 & 02:44:47.8  & -08:16:06.0
 \\

VST-ATLAS J025-33 & 6.31 & 2.18-6.31 & 18926 & 22.1 & 01:40:55.56 & -33:27:45.72
 \\

VIK J0305-3150 & 6.61 & 2.31-6.61 & 26400 & 7.8 & 03:05:16.916  & -31:50:55.98
 \\

BR0305-4957 & 4.78 & 1.95-4.78 & 2409 & 29.4 & 03:07:22.9  & -49:45:48.0
 \\

BR0322-2928 & 4.62 & 1.95-4.62 & 2409 & 21.1 & 03:24:44.3  & -29:18:21.1 
 \\

BR0331-1622 & 4.32 & 1.95-4.32 & 1944 & 15.1 & 03:34:13.4  & -16:12:05.2
 \\

SDSS J0331-0741 & 4.74 & 1.95-4.74 & 2177 & 6.2 & 03:31:19.7  & -07:41:43.1
 \\

SDSS J0332-0654 & 3.69 & 1.95-3.69 & 2409 & 5.6 & 03:32:23.5  & -06:54:50.0
 \\

SDSS J0338+0021 & 5.02 & 1.95-5.02 & 1817 & 4.3 & 03:38:29.3  & +00:21:56.5
 \\

SDSS J0344-0653 & 3.96 & 1.95-3.83 & 3022 & 6.6 & 03:44:02.85 & -06:53:00.6
 \\

BR0353-3820 & 4.58 & 1.95-4.58 & 1200 & 26.7 & 03:55:04.9  & -38:11:42.3
 \\

PSO J036+03 & 6.54 & 2.28-6.54 & 10240 & 12.3 & 02 26 01.88  & +03 02 59.4
 \\

BR0418-5723 & 4.37 & 1.95-4.37 & 4200 & 8.5 & 04:19:50.9  & -57:16:13.0
 \\

PSO J071-02 & 5.70 & 1.95-5.40 & 1817 & 6.9 & 04:45:48.18  & -02:19:59.8
 \\

DES J0454-4448 & 6.09 & 2.08-6.09 & 19878 & 12.3 & 04:54:01.79  & -44:48:31.1 
 \\

PSO 065-26 & 6.14 & 2.10-6.14 & 7228 & 11.0 & 04:21:38.05  & -26:57:15.6
 \\

PSO J071-02 & 5.69 & 1.95-5.40 & 3614 & 8.9 & 04:45:48.18 & -02:19:59.8
 \\

SDSS J0759+1800 & 4.79 & 1.95-4.79 & 2409 & 3.5 &  07:59:07.57 & +18:00:54.71
 \\

SDSS J0817+1351 & 4.39 & 1.95-4.39 & 2409 & 6.5 & 08:17:40.50 & +13:51:35.0
 \\

SDSS J0818+0719 & 4.58 & 1.95-4.39 & 2409 & 11.4 & 08:18:06.9  & +07:19:20.0
 \\

SDSS J0818+1722 & 6.02 & 2.00-5.40 & 9000 & 10.2 & 08:18:27.10  & +17:22:51.79
 \\

SDSS J0824+1302 & 5.19 & 1.95-5.19 & 4818 & 7.8 & 08:24:54.02  & +13:02:17.01
 \\

SDSS J0836+0054 & 5.81 & 1.96-5.40 & 33200 & 32.3 & 08:36:43.9  & +00:54:53.3
 \\

SDSS J0842+1218 & 6.07 & 2.07-6.07 & 7228 & 6.6 & 15:58:50.99  & -07:24:09.6
 \\

SDSS J0842+0637 & 3.66 & 1.95-3.66 & 2409 & 9.1 & 08:42:03.3  & +06:37:52.0
 \\

SDSS J0902+0851 & 5.23 & 1.95-5.20 & 3001 & 4.0 & 09:02:45.76  & +08:51:15.8
 \\

SDSS J0935+0022 & 3.75 & 1.96-5.40 & 1817 & 12.0 & 09:35:56.9  & +00:22:55.0
 \\

SDSS J0949+0335 & 4.05 & 1.95-3.83 & 1817 & 13.6 & 09:49:32.3  & +03:35:31.0
 \\

SDSS J1015+0020 & 4.40 & 1.95-4.40 & 3001 & 10.5 & 10:15:49.0  & +00:20:20.0
 \\

SDSS J1020+0922 & 3.64 & 1.95-3.64 & 2409 & 15.2 & 10:20:40.6  & +09:22:54.0
 \\

SDSS J1030+0524 & 6.31 & 2.18-6.31 & 14400 & 5.0 & 10 30 27.1  & +05 24 55.1 
 \\

SDSS J1037+0704 & 4.10 & 1.95-3.83 & 2726 & 8.8 & 10:37:32.4  & +07:04:26.0 
 \\

VIK J1048-0109 & 6.64 & 2.32-6.64 & 37578 & 5.4 & 10:48:19.08  & -01:09:40.3
 \\

SDSS J1100+1122 & 4.72 & 1.95-4.72 & 2409 & 9.6 & 11:00:45.23  & +11:22:39.14
 \\

SDSS J1101+0531 & 4.98 & 1.95-4.98 & 3001 & 5.0 & 11:01:34.4  & +05:31:33.0
 \\

SDSS J1110+0244 & 4.12 & 1.95-3.83 & 2409 & 18.6 & 11:10:08.6  & +02:44:58.0
 \\

SDSS J1115+0829 & 4.63 & 1.95-4.63 & 2409 & 8.1 & 11:15:23.2  & +08:29:18.0
 \\

ULAS J1120+0641 & 7.09 & 2.51-7.08 & 46243 & 11.4 & 11:20:01.48  & +06:41:24.3
 \\

SDSS J1132+1209 & 5.16 & 1.95-5.16 & 3001 & 8.9 & 11:32:46.50  & +12:09:01.69
 \\

SDSS J1135+0842 & 3.83 & 1.95-3.83 & 2409 & 17.7 & 11:35:36.4  & +08:42:19.0
 \\

ULAS J1148+0702 & 6.32 & 2.17-6.29 & 6023 & 6.2 & 11:48:03.29  & +07:02:08.3
 \\

PSO J183-12 & 5.86 & 1.98-5.40 & 19513 & 21.2 & 12:13:11.81  & -12:46:03.45
 \\

SDSS J1249-0159 & 3.64 & 1.95-3.64 & 1817 & 18.2 & 12:49:57.2  & -01:59:28.0 
 \\

SDSS J1253+1046 & 4.91 & 1.95-4.91 & 3001 & 9.4 & 12:53:53.35  & +10:46:03.19
 \\

SDSS J1257-0111 & 4.11 & 1.95-3.83 & 3001 & 21.9 & 12:57:59.2  & -01:11:30.0
 \\

SDSS J1305+0521 & 4.09 & 1.95-3.83 & 1363 & 8.8 & 13:05:02.3  & +05:21:51.0
 \\

SDSS J1306+0356 & 6.02 & 2.04-5.99 & 15682 & 9.2 & 13:06:08.3  & +03:56:26.3
 \\

ULAS 1319+0950 & 6.13 & 2.10-6.13 & 19275 & 5.0 & 13:19:11.3  & +09:50:51.
 \\

SDSS J1402+0146 & 4.16 & 1.95-3.83 & 1902 & 15.0 & 14:02:48.1  & +01:46:34.0
 \\

SDSS J1408+0205 & 4.01 & 1.95-3.83 & 2409 & 9.9 & 14:08:50.9  & +02:05:22.0
 \\

SDSS J1411+1217 & 5.90 & 2.01-5.93 & 3600 & 8.6 & 14:11:11.29  &   +12:17:37.40
 \\

PSO J213-22 & 5.92 & 2.00-5.40 & 18007 & 11.2 & 14:13:27.12  & -22:33:42.25
 \\

Q1422+2309 & 3.62 & 1.95-3.65 & 1226 & 47.2 & 14:24:38.09 & +22:56:00.6
\\

\enddata                     
\label{tab:qsoList}

\end{deluxetable*}

\setcounter{table}{0}
\begin{deluxetable*}{l c c c c c c c}[!htbp]
\tablecaption{FIRE \mgii Survey Sightlines}
\tablehead{ \colhead{Quasar}  & \colhead{$z_{em}$} & \colhead{$\Delta z$} & \colhead{$t_{exp}$ } & \colhead{Median SNR\tablenotemark{a}} & \colhead{RA} & \colhead{DEC} \\ \colhead{} & \colhead{} & \colhead{} & \colhead{(sec.)} & \colhead{(pixel$^{-1}$)}  }

\startdata

SDSS J1433+0227 & 4.72 & 1.95-4.72 & 2409 & 13.4 & 14:33:52.2 & +02:27:13.0
 \\

SDSS J1436+2132 & 5.25 & 1.95-5.24 & 2409 & 4.6 & 14:36:05.00 & +21:32:39.27
 \\

SDSS J1444-0101 & 4.51 & 1.95-4.51 & 2409 & 8.6 & 14:44:07.6 & -01:01:52.0
 \\

CFHQS 1509-1749 & 6.12 & 2.10-6.12 & 9900 & 17.6 & 15:09:41.78 & -17:49:26.80
 \\

SDSS J1511+0408 & 4.69 & 1.95-4.67 & 3001 & 11.6 & 15:11:56.0 & +04:08:02.0
 \\

SDSS J1532+2237 & 4.42 & 1.95-4.63 & 2409 & 14.3 & 15:32:47.41 & +22:37:04.18
 \\

SDSS J1538+0855 & 3.55 & 1.95-3.55 & 1363 & 24.2 & 15:38:30.5 & +08:55:17.0
 \\

PSO J159-02 & 6.38 & 2.20-6.35 & 6615 & 7.2 & 10:36:54.19 & -02:32:37.9
 \\

SDSS J1601+0435 & 3.85 & 1.95-3.83 & 3011 & 6.8 & 16:01:06.6 & +04:35:34.0
 \\

SDSS J1606+0850 & 4.55 & 1.95-4.55 & 2400 & 3.7 & 16:06:51.0 & +08:50:37.0
 \\

SDSS J1611+0844 & 4.53 & 1.95-4.53 & 4501 & 8.0 & 16:11:05.6 & +08:44:35.0
 \\

SDSS J1616+0501 & 4.88 & 1.95-4.88 & 3000 & 17.8 & 16:16:22.1 & +05:01:27.0
 \\

SDSS J1620+0020 & 4.09 & 1.95-3.83 & 972 & 7.0 & 16:20:48.7 & +00:20:05.0
 \\

SDSS J1621-0042 & 3.70 & 1.95-3.70 & 1204 & 26.1 & 16:21:16.9 & -00:42:50.0
 \\

SDSS J1626+2751 & 5.20 & 1.95-5.20 & 3614 & 16.6 &  16:26:26.50 & +27:51:32.4
 \\

PSO J167-13 & 6.51 & 2.26-6.51 & 19233 & 5.8 & 11:10:33.98 & -13:29:45.60
 \\

PSO J183+05\tablenotemark{b} & 6.45 & 2.24-6.45 & 11730 & 8.0 & ${\ast}$ & ${\ast}$
 \\

PSO J209-26 & 5.72 & 1.95-5.40 & 4818 & 11.1 & 13:56:49.41 & -26:42:30.23
 \\

SDSS J2147-0838 & 4.59 & 1.95-4.59 & 2409 & 13.8 & 21:47:25.7 & -08:38:34.0
 \\

PSO J217-16 & 6.14 & 2.10-6.14 & 22509 & 15.3 & 14:28:21.39 & -16:02:43.30
 \\

VIK J2211-3206 & 6.31 & 2.19-6.33 & 3001 & 6.9 & 22:11:12.391 & -32:06:12.95
 \\

SDSS J2228-0757\tablenotemark{b} & 5.14 & 1.95-5.14 & 3600 & 4.9 & $\ast$ & $\ast$
 \\

PSO J231-20 & 6.59 & 2.30-6.59 & 9637 & 9.7 & 15:26:37.84 & -20:50:00.7
 \\

SDSS J2310+1855 & 6.00 & 2.06-6.04 & 14400 & 17.5 & 23:10:38.89 & +18:55:19.93
 \\

VIK J2318-3113\tablenotemark{b} & 6.51 & 2.26-6.51 & 10504 & 4.3 & $\ast$ & ${\ast}$
 \\

BR2346-3729 & 4.21 & 1.95-3.83 & 2409 & 11.0 & 23:49:13.8 & -37:12:58.9
 \\

VIK J2348-3054 & 6.90 & 2.43-6.89 & 13822 & 4.8 & 23:48:33.34 & -30:54:10.24
 \\

PSO J239-07 & 6.11 & 2.09-6.11 & 12649 & 11.5 & 15:58:50.99 & -07:24:09.59
 \\

PSO J242-12 & 5.83 & 1.96-5.40 & 3001 & 6.4 & 16:09:45.53 & -12:58:54.11
 \\

PSO J247+24\tablenotemark{b} & 6.47 & 2.25-6.47 & 6626 & 4.0 & ${\ast}$ & ${\ast}$
 \\

PSO J308-27 & 5.80 & 1.95-5.40 & 12004 & 7.5 & 20:33:55.91 & -27:38:54.60

\enddata \tablenotetext{a}{Median signal-to-noise ratio per pixel
  across \mgii pathlength.}
  \tablenotetext{b}{Denotes survey quasars with unpublished coordinates, because discovery papers are in preparation \citep{mazzucchelli2017_inprep, venemans2018_inprep}}
\label{tab:qsoList}
\nonumber

\end{deluxetable*}

\section{Analysis}

We have used the software pipeline developed for the original survey
to conduct our analysis. The full details of this analysis along with
tests and development of the methodology are described in Paper
I. Below, we summarize the major steps and describe updates to the
process.

\subsection{Continuum fitting}
We fit an automatically-generated continuum to each flux-calibrated
spectrum via custom IDL routines.  These routines first generate an
initial mask of absorption features identified by pixel fluxes near
zero.  The masked spectrum is then split into segments of width 1250
km s$^{-1}$, which are median filtered to remove narrow absorption
features.  Each segment is allocated two knots for a cubic spline
interpolation fit to the continuum across the full spectrum.  The
knots' locations are determined from the statistics of the median
filtering process.  The spline fit was iterated between two and five
times with rejection of outlying pixels, to achieve convergence of the
fit.

\subsection{\mgii Line Finding}
We then searched the continuum-normalized data for cosmological
absorbers using a matched filter, composed of two Gaussians separated
by the intrinsic \mgii doublet spacing. An initial candidate catalog was constructed with redshifts located at SNR$>5$ peaks in the data/kernel cross-correlation. This was repeated for a set of Gaussian kernels with 
FWHM between 37.5 to 150 \kms ~to match systems of different intrinsic width. 
Since the matched filter returns many false positive detections (principally caused by OH sky line residuals), we subjected each candidate to a set of
consistency checks to eliminate obviously spurious systems.  These are
described in detail in Paper I; they are accomplished by explicitly fitting individual Gaussians to each component of the doublet, and then verifying from the fit parameters that (1) $W_{2796}>W_{2803}$ (within measurement errors), (2) the  FWHM exceeds FIRE's resolution element, but is not larger than 25 pixels
(313 km s$^{-1}$, chosen empirically to minimize BAL contamination and
continuum errors), (3) the amplitude of the Gaussian fit must be net positive and exceed the local noise RMS, (4) single systems cannot have broad kinematic
components separated by more than three times the total FWHM, and are instead split into separate absorbers in the sample. While criterion (3) might at first seem superfluous given the $5\sigma$ threshold for creating the parent catalog, it proves helpful in screening some narrow positive peaks in the filtered spectrum located immediately adjacent to narrow negative peaks (from sky lime residuals in crowded bandheads).  

Each \mgii candidate that survived this screening was visually
inspected, and the accepted systems were incorporated into the final
sample presented in Table 2 and plotted in Figure
\ref{fig:alldoublets}. We measured rest-frame equivalent widths and associated errors by direct summation of the unweighted spectral pixels (and quadrature summation of the error vector) rather than parameterized fits. Table 2 also reports a velocity width $\Delta v$ for each system corresponding to the interval on either side of the the line centroid where the normalized absorption profile remains below unity. 

\begin{figure*}[htp!]
	\begin{minipage}{\textwidth}
	\includegraphics[width=\textwidth]{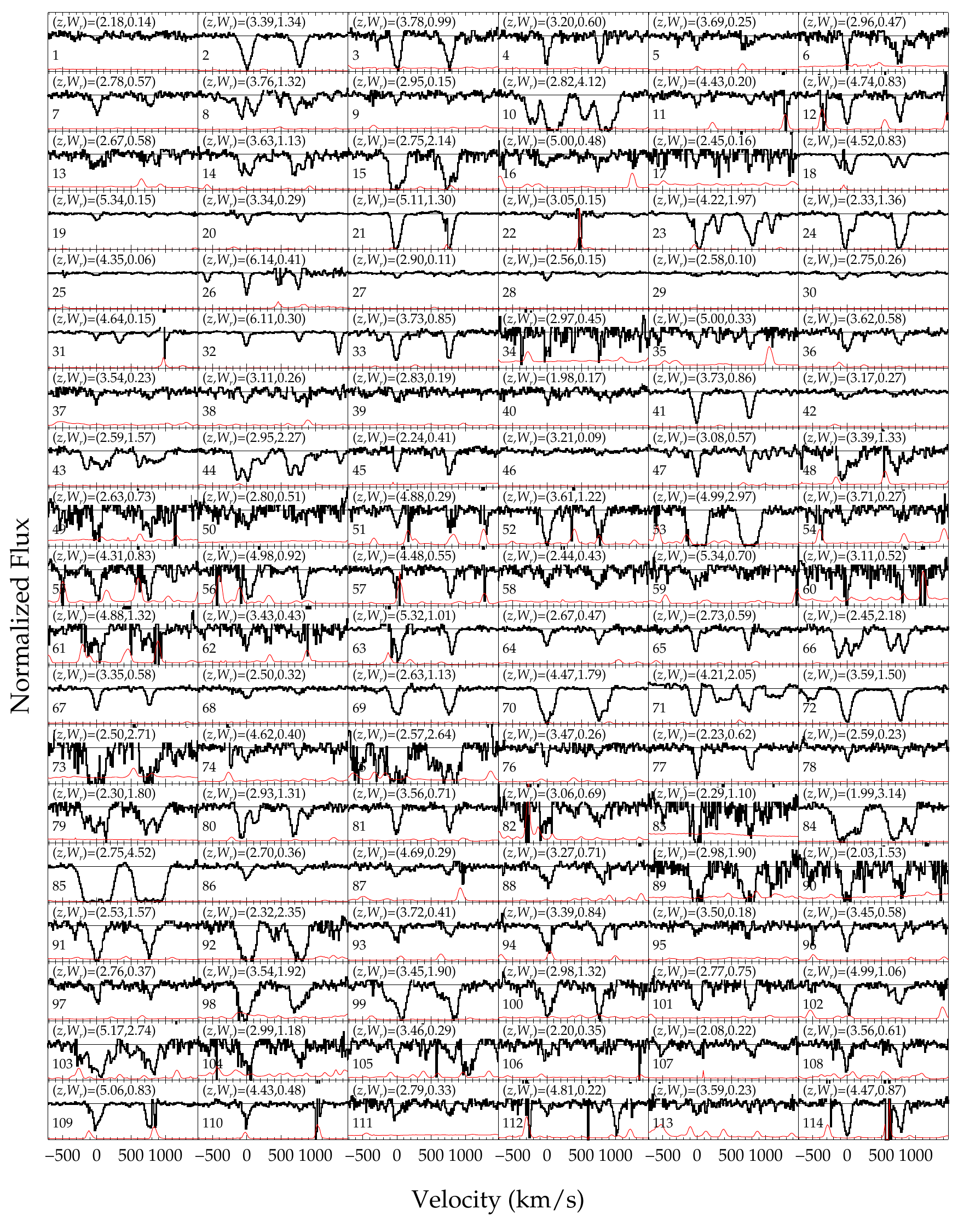}
	\end{minipage}
	\label{fig:alldoublets}
\end{figure*}

\begin{figure*}[htp!]
	\begin{minipage}{\textwidth}
	\includegraphics[width=\textwidth]{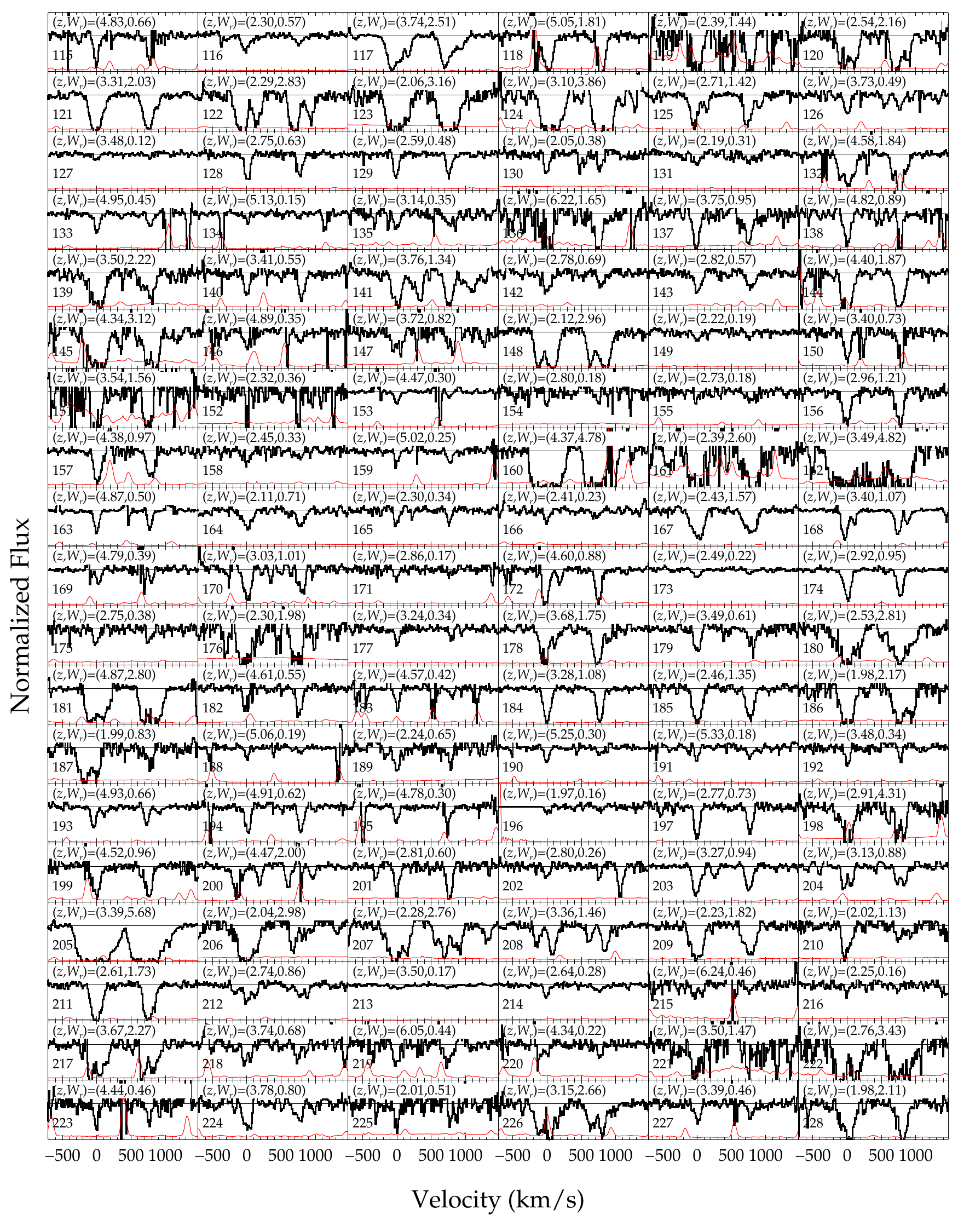}
	\end{minipage}
	\label{fig:alldoublets}
\end{figure*}

\begin{figure*}[htp!]
	\begin{minipage}{\textwidth}
	\includegraphics[width=\textwidth]{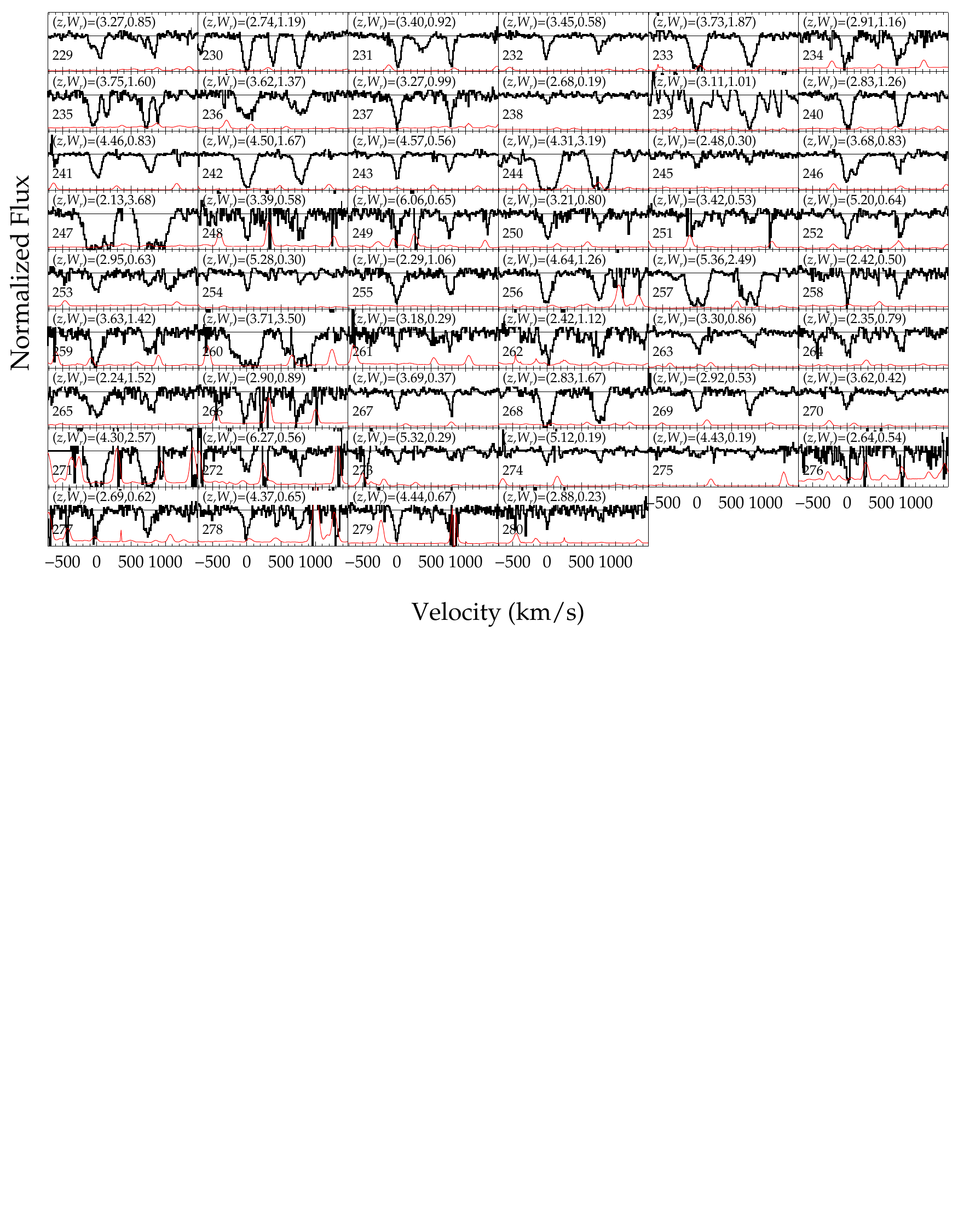}
	\end{minipage}
	\caption{Full \mgii doublet sample identified in our survey,
          plotted as continuum-normalized spectra against velocity separation from the $2796 \AA$
          transition. The spectra are plotted on a linear scale from zero (bottom of plot) with the thin black line in each panel denoting unity. All doublets are shown in the order presented
          in Table \ref{tab:dlist}, with index number at lower left of
          each panel corresponding to the rown number in the Table.
          The thin red line in each panel indicates the $1\sigma$
          error in normalized flux for each pixel.}
	\label{fig:alldoublets}
\end{figure*}

\begin{deluxetable*}{ c l c c c c c c}
\tablewidth{0pc}\tablecaption{Summary of Absorption Properties for the FIRE \mgii Sample}\tablehead{ \colhead{Index \#} & \colhead{Sightline} & \colhead{$z$} & \colhead{$W_{r}(2796)$} & \colhead{$\sigma(2796)$} & \colhead{$\Delta v$\tablenotemark{d}} \\ \colhead{} & \colhead{} & \colhead{} & \colhead{(\AA)} & \colhead{(\AA)} & \colhead{(\kms)} }\startdata

1~ & Q0000-26 & 2.1839 & 0.140 & 0.032 & 167.8 \\
2~ & Q0000-26 & 3.3900 & 1.340 & 0.029 & 232.3 \\
3~ & BR0004-6224 & 3.7765 & 0.988 & 0.054 & 193.8 \\
4~ & BR0004-6224 & 3.2037 & 0.601 & 0.043 & 114.6 \\
5~ & BR0004-6224 & 3.6946 & 0.250 & 0.047 & 154.7 \\
6~ & BR0004-6224 & 2.9598 & 0.470 & 0.086 & 167.8 \\
7~ & BR0016-3544 & 2.7825 & 0.566 & 0.034 & 180.8 \\
8~ & BR0016-3544 & 3.7571 & 1.321 & 0.038 & 384.5 \\
9~ & BR0016-3544 & 2.9485 & 0.146 & 0.031 & 86.7 \\
10~ & BR0016-3544 & 2.8184 & 4.116 & 0.059 & 686.1 \\
11~ & SDSS J0040-0915 & 4.4268 & 0.204 & 0.023 & 128.2 \\
12~ & SDSS J0040-0915 & 4.7396 & 0.831 & 0.027 & 154.7 \\
13~ & SDSS J0040-0915 & 2.6715 & 0.582 & 0.057 & 346.6 \\
14~ & SDSS J0042-1020 & 3.6297 & 1.129 & 0.031 & 308.7 \\
15~ & SDSS J0042-1020 & 2.7550 & 2.141 & 0.023 & 334.0 \\
16~ & SDSS J0054-0109 & 4.9975 & 0.476 & 0.063 & 283.3 \\
17~ & SDSS J0054-0109 & 2.4471 & 0.159 & 0.065 & 55.9 \\
18~ & SDSS J0100+28 & 4.5192 & 0.834 & 0.029 & 308.7 \\
19~ & SDSS J0100+28 & 5.3389 & 0.147 & 0.005 & 128.2 \\
20~ & SDSS J0100+28 & 3.3376 & 0.286 & 0.021 & 114.6 \\
21~ & SDSS J0100+28 & 5.1084 & 1.300 & 0.014 & 193.8 \\
22\tablenotemark{a} & SDSS J0100+28 & 3.0515 & 0.151 & 0.012 & 167.8 \\
23\tablenotemark{b} & SDSS J0100+28 & 4.2230 & 1.971 & 0.035 & 598.3 \\
24~ & SDSS J0100+28 & 2.3255 & 1.364 & 0.009 & 180.8 \\
25~ & SDSS J0100+28 & 4.3479 & 0.060 & 0.008 & 167.8 \\
26~ & SDSS J0100+28 & 6.1437 & 0.415 & 0.006 & 86.7 \\
27~ & SDSS J0100+28 & 2.9001 & 0.114 & 0.005 & 128.2 \\
28~ & SDSS J0100+28 & 2.5620 & 0.147 & 0.009 & 114.6 \\
29~ & SDSS J0100+28 & 2.5819 & 0.098 & 0.011 & 232.3 \\
30~ & SDSS J0100+28 & 2.7501 & 0.264 & 0.009 & 167.8 \\
31~ & SDSS J0100+28 & 4.6435 & 0.149 & 0.010 & 114.6 \\
32~ & SDSS J0100+28 & 6.1118 & 0.300 & 0.005 & 114.6 \\
33~ & SDSS J0106+0048 & 3.7290 & 0.854 & 0.016 & 154.7 \\
34~ & VIK J0109-3047 & 2.9695 & 0.454 & 0.074 & 100.8 \\
35~ & VIK J0109-3047 & 5.0011 & 0.335 & 0.042 & 128.2 \\
36~ & SDSS J0113-0935 & 3.6167 & 0.581 & 0.046 & 180.8 \\
37~ & SDSS J0113-0935 & 3.5446 & 0.231 & 0.039 & 128.2 \\
38\tablenotemark{a,c} & SDSS J0113-0935 & 3.1140 & 0.257 & 0.031 & 141.5 \\
39~ & SDSS J0113-0935 & 2.8252 & 0.188 & 0.029 & 114.6 \\
40\tablenotemark{c} & SDSS J0127-0045 & 1.9785 & 0.168 & 0.021 & 296.0 \\
41~ & SDSS J0127-0045 & 3.7282 & 0.863 & 0.014 & 167.8 \\
42~ & SDSS J0127-0045 & 3.1688 & 0.270 & 0.019 & 346.6 \\
43~ & SDSS J0127-0045 & 2.5881 & 1.568 & 0.027 & 535.5 \\
44~ & SDSS J0127-0045 & 2.9458 & 2.272 & 0.040 & 397.1 \\
45~ & SDSS J0140-0839 & 2.2408 & 0.415 & 0.027 & 141.5 \\
46~ & SDSS J0140-0839 & 3.2122 & 0.093 & 0.014 & 141.5 \\
47\tablenotemark{a} & SDSS J0140-0839 & 3.0815 & 0.565 & 0.021 & 114.6 \\
48~ & SDSS J0157-0106 & 3.3860 & 1.332 & 0.083 & 409.7 \\
49~ & SDSS J0157-0106 & 2.6311 & 0.734 & 0.079 & 206.7 \\
50~ & SDSS J0157-0106 & 2.7980 & 0.510 & 0.052 & 359.3 \\
51~ & PSO J029-29 & 4.8762 & 0.289 & 0.028 & 114.6 \\
52~ & PSO J029-29 & 3.6086 & 1.219 & 0.054 & 180.8 \\
53~ & PSO J029-29 & 4.9864 & 2.966 & 0.102 & 472.7 \\
54~ & ULAS J0203+0012 & 3.7110 & 0.267 & 0.045 & 154.7 \\
55\tablenotemark{a} & ULAS J0203+0012 & 4.3129 & 0.830 & 0.095 & 154.7 \\
56~ & ULAS J0203+0012 & 4.9770 & 0.916 & 0.105 & 193.8 \\
57~ & ULAS J0203+0012 & 4.4818 & 0.548 & 0.195 & 128.2 \\
58~ & SDSS J0216-0921 & 2.4363 & 0.433 & 0.056 & 232.3 \\
59~ & SDSS J0231-0728 & 5.3391 & 0.699 & 0.056 & 296.0 \\
60\tablenotemark{a} & SDSS J0231-0728 & 3.1113 & 0.518 & 0.052 & 86.7
\enddata
\label{tab:dlist}
\end{deluxetable*}

\setcounter{table}{1}
\begin{deluxetable*}{ c l c c c c c c}
\tablewidth{0pc}\tablecaption{Summary of Absorption Properties for the FIRE \mgii Sample (\emph{Continued})}\tablehead{ \colhead{Index \#} & \colhead{Sightline} & \colhead{$z$} & \colhead{$W_{r}(2796)$} & \colhead{$\sigma(2796)$} & \colhead{$\Delta v$\tablenotemark{d}} \\ \colhead{} & \colhead{} & \colhead{} & \colhead{(\AA)} & \colhead{(\AA)} & \colhead{(\kms)} }\startdata

61~ & SDSS J0231-0728 & 4.8840 & 1.322 & 0.133 & 409.7 \\
62~ & SDSS J0231-0728 & 3.4298 & 0.431 & 0.037 & 167.8 \\
63~ & ATLAS J025-33 & 5.3153 & 1.007 & 0.050 & 154.7 \\
64~ & ATLAS J025-33 & 2.6666 & 0.470 & 0.024 & 154.7 \\
65~ & ATLAS J025-33 & 2.7340 & 0.591 & 0.017 & 128.2 \\
66~ & ATLAS J025-33 & 2.4460 & 2.183 & 0.031 & 409.7 \\
67~ & BR0305-4957 & 3.3545 & 0.576 & 0.016 & 154.7 \\
68~ & BR0305-4957 & 2.5023 & 0.322 & 0.027 & 206.7 \\
69~ & BR0305-4957 & 2.6295 & 1.127 & 0.021 & 245.1 \\
70~ & BR0305-4957 & 4.4669 & 1.789 & 0.016 & 283.3 \\
71\tablenotemark{b,c} & BR0305-4957 & 4.2120 & 2.047 & 0.018 & 761.4 \\
72~ & BR0305-4957 & 3.5916 & 1.503 & 0.020 & 232.3 \\
73~ & VIK J0305-3150 & 2.4962 & 2.707 & 0.122 & 397.1 \\
74~ & VIK J0305-3150 & 4.6202 & 0.401 & 0.040 & 128.2 \\
75~ & VIK J0305-3150 & 2.5652 & 2.638 & 0.124 & 573.2 \\
76~ & VIK J0305-3150 & 3.4650 & 0.256 & 0.035 & 100.8 \\
77~ & BR0322-2928 & 2.2291 & 0.617 & 0.023 & 128.2 \\
78~ & BR0331-1622 & 2.5933 & 0.230 & 0.024 & 100.8 \\
79~ & BR0331-1622 & 2.2952 & 1.804 & 0.076 & 460.1 \\
80~ & BR0331-1622 & 2.9277 & 1.311 & 0.055 & 359.3 \\
81~ & BR0331-1622 & 3.5566 & 0.714 & 0.039 & 154.7 \\
82\tablenotemark{a} & SDSS J0332-0654 & 3.0618 & 0.686 & 0.113 & 245.1 \\
83~ & SDSS J0338+0021 & 2.2947 & 1.103 & 0.091 & 128.2 \\
84~ & BR0353-3820 & 1.9871 & 3.142 & 0.036 & 548.1 \\
85~ & BR0353-3820 & 2.7537 & 4.519 & 0.020 & 824.0 \\
86~ & BR0353-3820 & 2.6965 & 0.357 & 0.018 & 180.8 \\
87~ & PSO J036+03 & 4.6947 & 0.295 & 0.027 & 167.8 \\
88~ & PSO J036+03 & 3.2745 & 0.710 & 0.028 & 180.8 \\
89\tablenotemark{a} & BR0418-5723 & 2.9780 & 1.896 & 0.080 & 334.0 \\
90~ & BR0418-5723 & 2.0305 & 1.533 & 0.074 & 245.1 \\
91~ & DES0454-4448 & 2.5264 & 1.566 & 0.050 & 257.9 \\
92~ & DES0454-4448 & 2.3174 & 2.350 & 0.047 & 384.5 \\
93~ & DES0454-4448 & 3.7234 & 0.407 & 0.048 & 180.8 \\
94~ & DES0454-4448 & 3.3932 & 0.842 & 0.082 & 206.7 \\
95~ & DES0454-4448 & 3.5017 & 0.176 & 0.043 & 100.8 \\
96~ & DES0454-4448 & 3.4500 & 0.582 & 0.020 & 128.2 \\
97~ & DES0454-4448 & 2.7565 & 0.370 & 0.023 & 114.6 \\
98~ & PSO J065-26 & 3.5381 & 1.923 & 0.133 & 346.6 \\
99~ & PSO J065-26 & 3.4480 & 1.902 & 0.029 & 257.9 \\
100\tablenotemark{a} & PSO J065-26 & 2.9829 & 1.315 & 0.070 & 257.9 \\
101~ & PSO J071-02 & 2.7732 & 0.747 & 0.042 & 167.8 \\
102~ & PSO J071-02 & 4.9944 & 1.059 & 0.073 & 257.9 \\
103~ & PSO J071-02 & 5.1735 & 2.738 & 0.111 & 371.9 \\
104\tablenotemark{a} & SDSS J0817+1351 & 2.9946 & 1.185 & 0.102 & 232.3 \\
105~ & SDSS J0817+1351 & 3.4648 & 0.293 & 0.055 & 193.8 \\
106\tablenotemark{c} & SDSS J0818+0719 & 2.2049 & 0.353 & 0.047 & 193.8 \\
107\tablenotemark{c} & SDSS J0818+0719 & 2.0832 & 0.217 & 0.028 & 167.8 \\
108~ & SDSS J0818+1722 & 3.5629 & 0.607 & 0.078 & 128.2 \\
109~ & SDSS J0818+1722 & 5.0649 & 0.834 & 0.063 & 128.2 \\
110~ & SDSS J0818+1722 & 4.4309 & 0.478 & 0.053 & 180.8 \\
111~ & SDSS J0824+1302 & 2.7919 & 0.327 & 0.055 & 154.7 \\
112~ & SDSS J0824+1302 & 4.8110 & 0.224 & 0.035 & 100.8 \\
113~ & SDSS J0824+1302 & 3.5872 & 0.234 & 0.071 & 86.7 \\
114~ & SDSS J0824+1302 & 4.4716 & 0.866 & 0.027 & 167.8 \\
115~ & SDSS J0824+1302 & 4.8308 & 0.659 & 0.047 & 114.6 \\
116~ & SDSS J0836+0054 & 2.2990 & 0.565 & 0.022 & 232.3 \\
117~ & SDSS J0836+0054 & 3.7443 & 2.509 & 0.016 & 510.4 \\
118~ & SDSS J0842+1218 & 5.0481 & 1.813 & 0.146 & 245.1 \\
119~ & SDSS J0842+1218 & 2.3921 & 1.437 & 0.251 & 193.8 \\
120~ & SDSS J0842+1218 & 2.5397 & 2.157 & 0.098 & 384.5
\enddata
\tablenotetext{a}{Poor telluric region}
\tablenotetext{b}{Missed by automated search algorithm}
\tablenotetext{c}{Not identified in Paper I}
\tablenotetext{d}{$\Delta v$ is defined as the total velocity interval about each line centroid within which the absorption profile remains below the fitted continuum.}
\nonumber
\label{}
\end{deluxetable*}

\setcounter{table}{1}
\begin{deluxetable*}{ c l c c c c c c}
\tablewidth{0pc}\tablecaption{Summary of Absorption Properties for the FIRE \mgii Sample (\emph{Continued})}\tablehead{ \colhead{Index \#} & \colhead{Sightline} & \colhead{$z$} & \colhead{$W_{r}(2796)$} & \colhead{$\sigma(2796)$} & \colhead{$\Delta v$\tablenotemark{d}} \\ \colhead{} & \colhead{} & \colhead{} & \colhead{(\AA)} & \colhead{(\AA)} & \colhead{(\kms)} }\startdata

121~ & SDSS J0949+0335 & 3.3105 & 2.026 & 0.044 & 296.0 \\
122~ & SDSS J0949+0335 & 2.2888 & 2.834 & 0.065 & 472.7 \\
123~ & SDSS J1015+0020 & 2.0588 & 3.161 & 0.133 & 510.4 \\
124\tablenotemark{b} & SDSS J1015+0020 & 3.1040 & 3.862 & 0.072 & 773.9 \\
125~ & SDSS J1015+0020 & 2.7103 & 1.417 & 0.073 & 296.0 \\
126~ & SDSS J1015+0020 & 3.7299 & 0.489 & 0.029 & 141.5 \\
127~ & SDSS J1020+0922 & 3.4786 & 0.117 & 0.016 & 128.2 \\
128~ & SDSS J1020+0922 & 2.7485 & 0.635 & 0.024 & 141.5 \\
129~ & SDSS J1020+0922 & 2.5933 & 0.482 & 0.027 & 128.2 \\
130~ & SDSS J1020+0922 & 2.0461 & 0.381 & 0.046 & 114.6 \\
131~ & SDSS J1030+0524 & 2.1881 & 0.315 & 0.021 & 371.9 \\
132~ & SDSS J1030+0524 & 4.5836 & 1.839 & 0.033 & 321.3 \\
133~ & SDSS J1030+0524 & 4.9481 & 0.455 & 0.023 & 141.5 \\
134~ & SDSS J1030+0524 & 5.1307 & 0.146 & 0.013 & 55.9 \\
135\tablenotemark{a} & SDSS J1037+0704 & 3.1373 & 0.349 & 0.062 & 193.8 \\
136~ & J1048-0109 & 6.2215 & 1.647 & 0.163 & 232.3 \\
137~ & J1048-0109 & 3.7465 & 0.952 & 0.061 & 167.8 \\
138~ & J1048-0109 & 4.8206 & 0.890 & 0.037 & 154.7 \\
139~ & J1048-0109 & 3.4968 & 2.221 & 0.076 & 434.9 \\
140~ & J1048-0109 & 3.4133 & 0.547 & 0.031 & 167.8 \\
141~ & SDSS J1100+1122 & 3.7566 & 1.342 & 0.055 & 232.3 \\
142~ & SDSS J1100+1122 & 2.7825 & 0.691 & 0.054 & 193.8 \\
143~ & SDSS J1100+1122 & 2.8225 & 0.570 & 0.063 & 206.7 \\
144~ & SDSS J1100+1122 & 4.3959 & 1.866 & 0.101 & 257.9 \\
145\tablenotemark{a} & SDSS J1101+0531 & 4.3431 & 3.118 & 0.264 & 460.1 \\
146~ & SDSS J1101+0531 & 4.8902 & 0.346 & 0.074 & 154.7 \\
147~ & SDSS J1101+0531 & 3.7191 & 0.820 & 0.063 & 257.9 \\
148~ & SDSS J1110+0244 & 2.1188 & 2.957 & 0.043 & 460.1 \\
149~ & SDSS J1110+0244 & 2.2232 & 0.193 & 0.024 & 141.5 \\
150~ & SDSS J1115+0829 & 3.4045 & 0.731 & 0.034 & 154.7 \\
151~ & SDSS J1115+0829 & 3.5427 & 1.557 & 0.172 & 219.5 \\
152~ & SDSS J1115+0829 & 2.3209 & 0.359 & 0.037 & 55.9 \\
153~ & ULAS J1120+0641 & 4.4725 & 0.298 & 0.015 & 128.2 \\
154~ & ULAS J1120+0641 & 2.8004 & 0.178 & 0.041 & 71.9 \\
155~ & SDSS J1132+1209 & 2.7334 & 0.180 & 0.031 & 206.7 \\
156~ & SDSS J1132+1209 & 2.9568 & 1.210 & 0.072 & 206.7 \\
157~ & SDSS J1132+1209 & 4.3801 & 0.968 & 0.098 & 193.8 \\
158~ & SDSS J1132+1209 & 2.4541 & 0.333 & 0.049 & 180.8 \\
159~ & SDSS J1132+1209 & 5.0162 & 0.249 & 0.027 & 114.6 \\
160~ & ULAS J1148+0702 & 4.3673 & 4.784 & 0.112 & 371.9 \\
161~ & ULAS J1148+0702 & 2.3858 & 2.600 & 0.287 & 359.3 \\
162~ & ULAS J1148+0702 & 3.4936 & 4.822 & 0.194 & 899.2 \\
163~ & PSO J183-12 & 4.8709 & 0.503 & 0.019 & 114.6 \\
164~ & PSO J183-12 & 2.1068 & 0.710 & 0.024 & 245.1 \\
165~ & PSO J183-12 & 2.2972 & 0.341 & 0.021 & 100.8 \\
166~ & PSO J183-12 & 2.4058 & 0.225 & 0.030 & 128.2 \\
167~ & PSO J183-12 & 2.4308 & 1.574 & 0.023 & 346.6 \\
168~ & PSO J183-12 & 3.3956 & 1.069 & 0.032 & 283.3 \\
169~ & SDSS J1253+1046 & 4.7930 & 0.394 & 0.052 & 100.8 \\
170\tablenotemark{a} & SDSS J1253+1046 & 3.0282 & 1.010 & 0.037 & 193.8 \\
171~ & SDSS J1253+1046 & 2.8565 & 0.169 & 0.030 & 100.8 \\
172~ & SDSS J1253+1046 & 4.6004 & 0.882 & 0.108 & 154.7 \\
173~ & SDSS J1257-0111 & 2.4894 & 0.223 & 0.019 & 154.7 \\
174~ & SDSS J1257-0111 & 2.9181 & 0.955 & 0.020 & 180.8 \\
175~ & SDSS J1305+0521 & 2.7527 & 0.375 & 0.040 & 128.2 \\
176~ & SDSS J1305+0521 & 2.3023 & 1.976 & 0.122 & 346.6 \\
177~ & SDSS J1305+0521 & 3.2354 & 0.337 & 0.026 & 128.2 \\
178~ & SDSS J1305+0521 & 3.6799 & 1.749 & 0.069 & 270.6 \\
179~ & SDSS J1306+0356 & 3.4898 & 0.607 & 0.033 & 167.8 \\
180~ & SDSS J1306+0356 & 2.5328 & 2.813 & 0.115 & 535.5
\enddata
\tablenotetext{a}{Poor telluric region}
\tablenotetext{b}{Missed by automated search algorithm}
\tablenotetext{c}{Not identified in Paper I}
\tablenotetext{d}{$\Delta v$ is defined as the total velocity interval about each line centroid within which the absorption profile remains below the fitted continuum.}
\nonumber
\label{}
\end{deluxetable*}

\setcounter{table}{1}
\begin{deluxetable*}{ c l c c c c c c}
\tablewidth{0pc}\tablecaption{Summary of Absorption Properties for the FIRE \mgii Sample (\emph{Continued})}\tablehead{ \colhead{Index \#} & \colhead{Sightline} & \colhead{$z$} & \colhead{$W_{r}(2796)$} & \colhead{$\sigma(2796)$} & \colhead{$\Delta v$\tablenotemark{d}} \\ \colhead{} & \colhead{} & \colhead{} & \colhead{(\AA)} & \colhead{(\AA)} & \colhead{(\kms)} }\startdata

181~ & SDSS J1306+0356 & 4.8651 & 2.804 & 0.068 & 180.8 \\
182~ & SDSS J1306+0356 & 4.6147 & 0.547 & 0.089 & 128.2 \\
183~ & ULAS J1319+0950 & 4.5681 & 0.420 & 0.062 & 128.2 \\
184~ & SDSS J1402+0146 & 3.2772 & 1.085 & 0.021 & 180.8 \\
185~ & SDSS J1408+0205 & 2.4622 & 1.349 & 0.047 & 219.5 \\
186~ & SDSS J1408+0205 & 1.9816 & 2.174 & 0.063 & 334.0 \\
187~ & SDSS J1408+0205 & 1.9910 & 0.830 & 0.038 & 219.5 \\
188~ & SDSS J1411+1217 & 5.0552 & 0.193 & 0.016 & 86.7 \\
189~ & SDSS J1411+1217 & 2.2367 & 0.647 & 0.040 & 193.8 \\
190~ & SDSS J1411+1217 & 5.2501 & 0.295 & 0.015 & 128.2 \\
191~ & SDSS J1411+1217 & 5.3315 & 0.182 & 0.016 & 100.8 \\
192~ & SDSS J1411+1217 & 3.4773 & 0.343 & 0.020 & 86.7 \\
193~ & SDSS J1411+1217 & 4.9285 & 0.659 & 0.024 & 128.2 \\
194~ & PSO J213-02 & 4.9125 & 0.623 & 0.030 & 128.2 \\
195~ & PSO J213-02 & 4.7777 & 0.295 & 0.028 & 114.6 \\
196~ & Q1422+2309 & 1.9720 & 0.163 & 0.020 & 128.2 \\
197~ & SDSS J1433+0227 & 2.7717 & 0.726 & 0.018 & 128.2 \\
198\tablenotemark{b} & SDSS J1436+2132 & 2.9070 & 4.309 & 0.030 & 610.9 \\
199~ & SDSS J1436+2132 & 4.5211 & 0.964 & 0.166 & 193.8 \\
200\tablenotemark{b} & SDSS J1444-0101 & 4.4690 & 2.002 & 0.173 & 472.7 \\
201~ & SDSS J1444-0101 & 2.8103 & 0.599 & 0.059 & 141.5 \\
202~ & SDSS J1444-0101 & 2.7967 & 0.264 & 0.044 & 114.6 \\
203~ & CFQS1509-1749 & 3.2662 & 0.940 & 0.018 & 180.8 \\
204\tablenotemark{a} & CFQS1509-1749 & 3.1272 & 0.878 & 0.076 & 245.1 \\
205~ & CFQS1509-1749 & 3.3925 & 5.679 & 0.056 & 811.5 \\
206~ & SDSS J1511+0408 & 2.0394 & 2.978 & 0.090 & 359.3 \\
207~ & SDSS J1511+0408 & 2.2771 & 2.756 & 0.081 & 485.3 \\
208~ & SDSS J1511+0408 & 3.3588 & 1.464 & 0.067 & 397.1 \\
209~ & SDSS J1511+0408 & 2.2310 & 1.825 & 0.051 & 321.3 \\
210~ & SDSS J1511+0408 & 2.0230 & 1.129 & 0.053 & 232.3 \\
211~ & SDSS J1532+2237 & 2.6116 & 1.725 & 0.032 & 245.1 \\
212~ & SDSS J1532+2237 & 2.7414 & 0.862 & 0.033 & 283.3 \\
213~ & SDSS J1538+0855 & 3.4979 & 0.165 & 0.012 & 346.6 \\
214~ & SDSS J1538+0855 & 2.6383 & 0.282 & 0.027 & 154.7 \\
215~ & PSO J159-02 & 6.2376 & 0.458 & 0.045 & 257.9 \\
216~ & PSO J159-02 & 2.2465 & 0.163 & 0.027 & 71.9 \\
217~ & PSO J159-02 & 3.6695 & 2.269 & 0.115 & 460.1 \\
218~ & PSO J159-02 & 3.7422 & 0.681 & 0.047 & 257.9 \\
219~ & PSO J159-02 & 6.0549 & 0.436 & 0.065 & 167.8 \\
220\tablenotemark{a} & PSO J159-02 & 4.3426 & 0.222 & 0.046 & 141.5 \\
221~ & SDSS J1601+0435 & 3.5007 & 1.467 & 0.129 & 308.7 \\
222~ & SDSS J1606+0850 & 2.7636 & 3.433 & 0.128 & 548.1 \\
223~ & SDSS J1606+0850 & 4.4426 & 0.464 & 0.041 & 100.8 \\
224~ & SDSS J1611+0844 & 3.7767 & 0.801 & 0.053 & 206.7 \\
225~ & SDSS J1611+0844 & 2.0144 & 0.506 & 0.059 & 100.8 \\
226\tablenotemark{a} & SDSS J1611+0844 & 3.1454 & 2.662 & 0.213 & 422.3 \\
227~ & SDSS J1611+0844 & 3.3861 & 0.464 & 0.038 & 141.5 \\
228\tablenotemark{c}~ & SDSS J1616+0501 & 1.9809 & 2.115 & 0.050 & 270.6 \\
229~ & SDSS J1616+0501 & 3.2747 & 0.853 & 0.021 & 180.8 \\
230~ & SDSS J1616+0501 & 2.7409 & 1.188 & 0.026 & 193.8 \\
231~ & SDSS J1616+0501 & 3.3955 & 0.916 & 0.055 & 141.5 \\
232~ & SDSS J1616+0501 & 3.4507 & 0.584 & 0.017 & 128.2 \\
233~ & SDSS J1616+0501 & 3.7327 & 1.866 & 0.057 & 321.3 \\
234~ & SDSS J1620+0020 & 2.9106 & 1.159 & 0.055 & 270.6 \\
235~ & SDSS J1620+0020 & 3.7515 & 1.601 & 0.070 & 232.3 \\
236~ & SDSS J1620+0020 & 3.6200 & 1.366 & 0.066 & 397.1 \\
237~ & SDSS J1620+0020 & 3.2726 & 0.988 & 0.047 & 167.8 \\
238~ & SDSS J1621-0042 & 2.6780 & 0.189 & 0.019 & 100.8 \\
239\tablenotemark{a} & SDSS J1621-0042 & 3.1057 & 1.013 & 0.013 & 232.3 \\
240~ & SDSS J1626+2751 & 2.8288 & 1.260 & 0.041 & 206.7
\enddata
\tablenotetext{a}{Poor telluric region}
\tablenotetext{b}{Missed by automated search algorithm}
\tablenotetext{c}{Not identified in Paper I}
\tablenotetext{d}{$\Delta v$ is defined as the total velocity interval about each line centroid within which the absorption profile remains below the fitted continuum.}
\nonumber
\label{}
\end{deluxetable*}

\setcounter{table}{1}
\begin{deluxetable*}{ c l c c c c c c}
\tablewidth{0pc}\tablecaption{Summary of Absorption Properties for the FIRE \mgii Sample (\emph{Continued})}\tablehead{ \colhead{Index \#} & \colhead{Sightline} & \colhead{$z$} & \colhead{$W_{r}(2796)$} & \colhead{$\sigma(2796)$} & \colhead{$\Delta v$\tablenotemark{d}} \\ \colhead{} & \colhead{} & \colhead{} & \colhead{(\AA)} & \colhead{(\AA)} & \colhead{(\kms)} }\startdata

241~ & SDSS J1626+2751 & 4.4619 & 0.829 & 0.014 & 219.5 \\
242~ & SDSS J1626+2751 & 4.4968 & 1.673 & 0.019 & 283.3 \\
243~ & SDSS J1626+2751 & 4.5682 & 0.561 & 0.025 & 128.2 \\
244\tablenotemark{a} & SDSS J1626+2751 & 4.3108 & 3.188 & 0.050 & 434.9 \\
245~ & SDSS J1626+2751 & 2.4822 & 0.300 & 0.032 & 141.5 \\
246~ & SDSS J1626+2751 & 3.6826 & 0.833 & 0.011 & 167.8 \\
247~ & SDSS J1626+2751 & 2.1320 & 3.679 & 0.091 & 321.3 \\
248~ & PSO J167-13 & 3.3889 & 0.581 & 0.036 & 141.5 \\
249~ & PSO J183+05 & 6.0643 & 0.653 & 0.096 & 141.5 \\
250~ & PSO J183+05 & 3.2071 & 0.803 & 0.042 & 180.8 \\
251~ & PSO J183+05 & 3.4184 & 0.533 & 0.077 & 219.5 \\
252~ & PSO J209-26 & 5.2021 & 0.643 & 0.025 & 154.7 \\
253~ & PSO J209-26 & 2.9505 & 0.631 & 0.048 & 206.7 \\
254~ & PSO J209-26 & 5.2758 & 0.299 & 0.020 & 100.8 \\
255~ & SDSS J2147-0838 & 2.2863 & 1.058 & 0.049 & 206.7 \\
256~ & PSO J217-16 & 4.6420 & 1.261 & 0.044 & 219.5 \\
257\tablenotemark{a} & PSO J217-16 & 5.3571 & 2.489 & 0.029 & 359.3 \\
258~ & PSO J217-16 & 2.4166 & 0.501 & 0.050 & 128.2 \\
259~ & VIK J2211-3206 & 3.6302 & 1.416 & 0.092 & 257.9 \\
260~ & VIK J2211-3206 & 3.7144 & 3.505 & 0.068 & 623.4 \\
261~ & SDSS J2228-0757 & 3.1754 & 0.287 & 0.038 & 71.9 \\
262~ & PSO J231-20 & 2.4191 & 1.115 & 0.090 & 257.9 \\
263~ & SDSS J2310+1855 & 3.2998 & 0.856 & 0.058 & 257.9 \\
264~ & SDSS J2310+1855 & 2.3510 & 0.789 & 0.052 & 193.8 \\
265~ & SDSS J2310+1855 & 2.2430 & 1.523 & 0.068 & 334.0 \\
266~ & VIK J2318-3113 & 2.9030 & 0.887 & 0.075 & 219.5 \\
267~ & BR2346-3729 & 3.6922 & 0.371 & 0.019 & 128.2 \\
268~ & BR2346-3729 & 2.8300 & 1.665 & 0.054 & 270.6 \\
269~ & BR2346-3729 & 2.9226 & 0.535 & 0.041 & 167.8 \\
270~ & BR2346-3729 & 3.6188 & 0.422 & 0.036 & 141.5 \\
271\tablenotemark{a} & VIK J2348-3054 & 4.2996 & 2.567 & 0.118 & 384.5 \\
272~ & VIK J2348-3054 & 6.2682 & 0.564 & 0.062 & 167.8 \\
273~ & PSO J239-07 & 5.3238 & 0.287 & 0.024 & 141.5 \\
274~ & PSO J239-07 & 5.1209 & 0.193 & 0.022 & 114.6 \\
275~ & PSO J239-07 & 4.4276 & 0.193 & 0.018 & 141.5 \\
276~ & PSO J242-12 & 2.6351 & 0.543 & 0.075 & 114.6 \\
277~ & PSO J242-12 & 2.6880 & 0.620 & 0.087 & 180.8 \\
278~ & PSO J242-12 & 4.3658 & 0.646 & 0.086 & 154.7 \\
279~ & PSO J242-12 & 4.4351 & 0.671 & 0.041 & 141.5 \\
280~ & PSO J308-27 & 2.8797 & 0.229 & 0.032 & 71.9 \\
\hline
\\
\nodata~ & BR0004-6224 & 2.663 & 0.260 & 0.045 & 58.0 \\
\nodata~ & BR0004-6224 & 2.908 & 0.596 & 0.047 & 83.3 \\
\nodata~ & SDSS J1030+0525 & 2.780 & 2.617 & 0.069 & 583.9 \\
\nodata~ & SDSS J1306+0356 & 4.882 & 1.941 & 0.079 & 248.8 \\
\nodata~ & SDSS J1402+0146 & 3.454 & 0.341 & 0.016 & 173.3 \\
\nodata~ & Q1422+2309 & 3.540 & 0.169 & 0.011 & 130.0 \\
\nodata~ & SDSS J2310+1855 & 2.243 & 1.441 & 0.050 & 292.1
\enddata
\tablenotetext{a}{Poor telluric region}
\tablenotetext{b}{Missed by automated search algorithm}
\tablenotetext{c}{Not identified in Paper I}
\tablenotetext{d}{$\Delta v$ is defined as the total velocity interval about each line centroid within which the absorption profile remains below the fitted continuum.}
\nonumber
\label{}
\end{deluxetable*}

\begin{deluxetable*}{ c l c c c c c c}
\tablewidth{0pc}\tablecaption{Proximate \mgii Systems}\tablehead{ \colhead{Index \#} & \colhead{Sightline} & \colhead{$z$} & \colhead{$W_{r}(2796)$} & \colhead{$\sigma(2796)$} & \colhead{$\Delta v$} \\ \colhead{} & \colhead{} & \colhead{} & \colhead{(\AA)} & \colhead{(\AA)} & \colhead{(\kms)} }\startdata

1 ~&  PSO J065-26  &  6.122  &  2.346  &  0.038  &  553.4 \\
2 ~&  SDSS J0140-0839  &  3.703  &  0.584  &  0.015  &  216.0 \\
3 ~&  SDSS J1436+2132  &  4.522  &  0.973  &  0.189  &  332.8 \\
4 ~&  SDSS J1626+2751  &  5.178  &  1.416  &  0.022  &  518.6 \\
5 ~&  PSO J183+05  &  6.404  &  0.774  &  0.053  &  356.6

\enddata
\nonumber
\label{tab:prox_mgii}
\end{deluxetable*}

For consistency, we have redone the line finding for the sightlines
presented in Paper I. A complete list of these doublets and their
continuum-normalized profiles are included in Table~\ref{tab:dlist}
and Figure~\ref{fig:alldoublets}, respectively. Differences in user
acceptances/rejections are noted in the table: in general, as the
visual inspection step was carried out by a different user than in the
original survey (SC and MM, respectively), we tended to be more
optimistic in accepting borderline candidates for \mgii
doublets. These tendencies are reflected in the user-rating
calibration, discussed below. In addition, we serendipitously
identified five systems excluded by the automated search algorithm.
These are reported and flagged in the table of absorbers, but they are
omitted from calculations of the \mgii population statistics, because
the statistical calculations account for such missed systems via
incompleness simulations. In the process of the visual identification,
we also identified five \mgii absorbers which were not included in our
sample due to their proximity to the background quasar; these are
listed with their associated properties in Table~\ref{tab:prox_mgii}.
The proximate absorbers in the two PS1 quasars are of particular
interest and will be discussed in detail in forthcoming work
\citep{banados_inprep}.

\subsection{Automated Completeness Test}
We ran a large Monte Carlo simulation to quantify the the completeness
of the automated line-finding algorithm. For each QSO, 10,000
simulated \mgii doublets with equivalent widths uniformly distributed
between 0.05 and 0.95 \AA ~and random redshifts were injected into the
spectrum (from which the real doublets were previously removed and
replaced with noise) and then subjected to the automated line-finding
algorithm. The rates at which these simulated doublets were recovered
were then binned into an automated completeness grid by redshift and
equivalent width (with $dz = 0.02$ and $dW = 0.01$ \AA) for each QSO,
which we will call $L_q(z,W)$. These computationally intensive
simulations were run on the {\tt antares} computing cluster at the MIT
Kavli Institute.
  
\subsection{User-Rating Calibration}
A subset of the automatically simulated doublets were inspected
visually to evaluate the efficacy of the human inspection step in our
doublet-finding procedure. In particular, the user may either reject a
real \mgii system or accept a false positive, thus requiring a
correction to our statistical calculations. We inspected 1000 such
simulated doublets, with the important difference that the user-test
systems had a slightly larger velocity spacing than legitimate \mgii
doublets.  This ensures that any ``doublets'' identified by the
machine are either artifically injected (and should therefore be
accepted) or correlated noise (and should be rejected).

While inspecting these false-spacing doublets, we identified three
very large absorbers, likely not due to \mgiinsp. These were manually
excised and masked from our Monte Carlo data so that only injected
doublets and correlated noise factored into the user ratings
calculation.  The user then either accepted or rejected the remaining
candidate doublets, and the success rates at which the user identified
real systems and rejected false positives were used to calculate a
total completeness for each QSO.
  
\begin{figure}[b]
	\begin{minipage}{0.5\textwidth}
	\includegraphics[width=\textwidth]{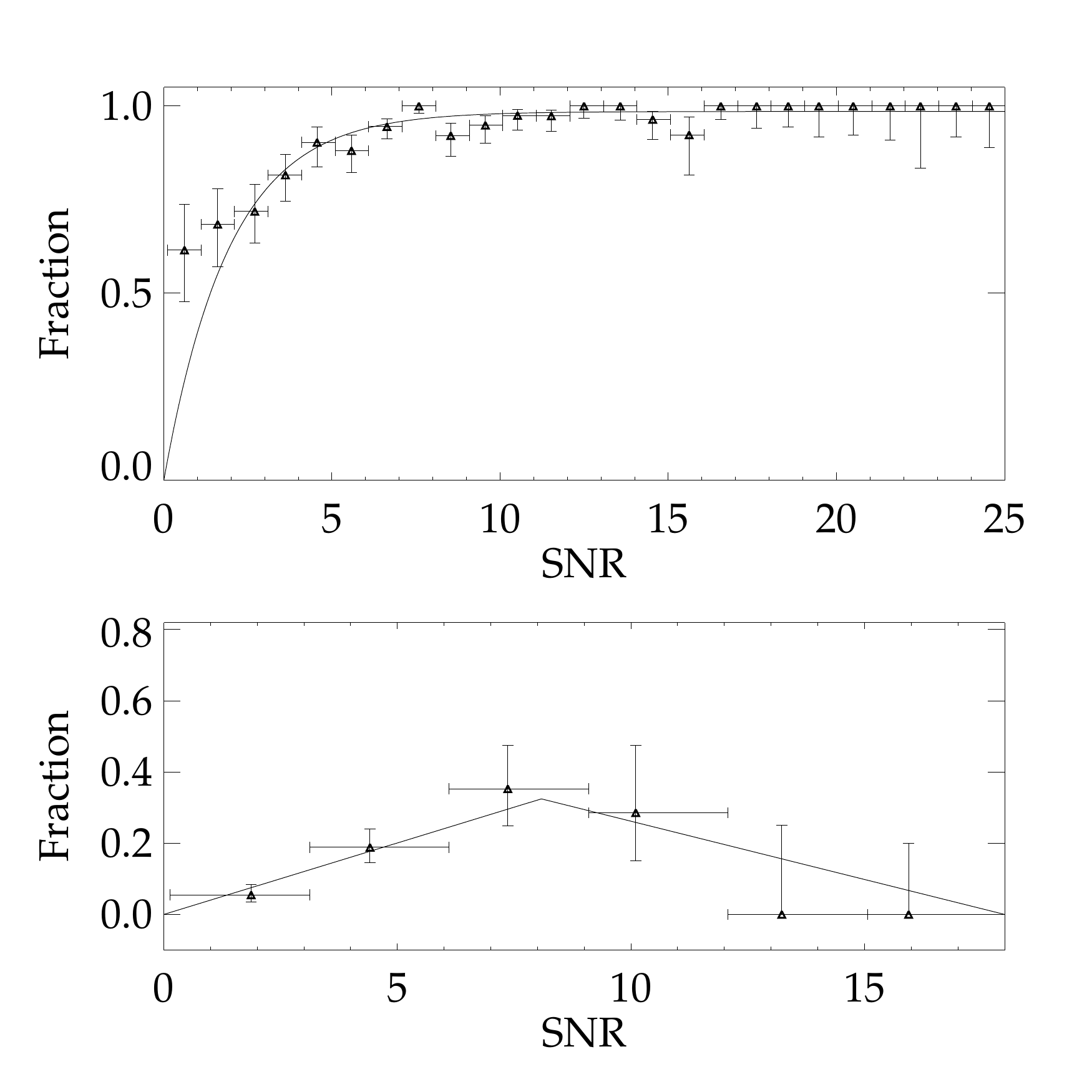}
	\end{minipage}
	\caption{The user acceptance rates for real injected systems
          (top) and false positives (bottom), binned by the detection SNR for each doublet (i.e. the significance of the peak in cross-correlation between data and matched filter). The error
          bars represent the Wilson score interval, and the lines are
          the MLE fits for these rates.  In the bottom panel, the errors are larger at high SNR because fewer injected systems are detected at SNR>10 in the presence of noise.}
	\label{fig:user-rates}
\end{figure}

As discussed in Paper I, the time-consuming nature of visual
inspection precludes the use of finely grained bins in $W_r$ and $z$,
but we found that the acceptance rate for real systems and false
positives depended primarily on the SNR of the candidate
doublets. They can be parametrized with SNR as follows:
  
\begin{equation}
	P^{\rm Mg II}(s) = P_{\infty}(1-e^{-s/s_c})
\end{equation}
\begin{equation}
	P^{\rm FP}(s) = 
	\begin{cases}
	P^{\rm FP}_{\rm max} (\frac{s}{s_p}), \quad s \leq s_p \\
	P^{\rm FP}_{\rm max} (\frac{s-s_f}{s_p-s_f}), \quad s > s_p
	\end{cases},
\end{equation}

where $P^{\rm Mg II}$ and $P^{\rm FP}$ are the acceptance rates for real
systems and false positives, respectively, and $P_{\infty}$, $s_c$,
$P^{\rm FP}_{\rm max}$, $s_p,$ and $s_f$ are free parameters fit by
maximum-likelihood estimation (MLE). Plots of the user acceptance
rates are given in Figure~\ref{fig:user-rates}.  Comparing the ratings
of SC and MM it is apparent that SC correctly identified a higher
fraction of \mgii doublets at low SNR, but this comes at the expense
of a higher false-positive rate.  After proper calibration these
tendencies should cancel, and indeed we will find very similar
statistical results as Paper I in areas where both may be
compared.

For each individual QSO, the user-acceptance rates were then estimated
as functions of the equivalent width $W_r$ to give $A^{\rm Mg II}(W,z)$
and $A^{\rm FP}(W,z)$, the acceptance rates for real systems and false
positives, respectively. The total completeness fraction for each QSO
$q$, $C_q(z,W)$, was then calulated as the product of the automated
completeness fraction and the user acceptance rate, namely

\begin{equation}
C_q(z,W) = A^{\rm MgII}(z,W) L_q(z,W).
\end{equation}

\begin{figure}[t]
	\begin{minipage}{0.5\textwidth}
	\includegraphics[width=\textwidth]{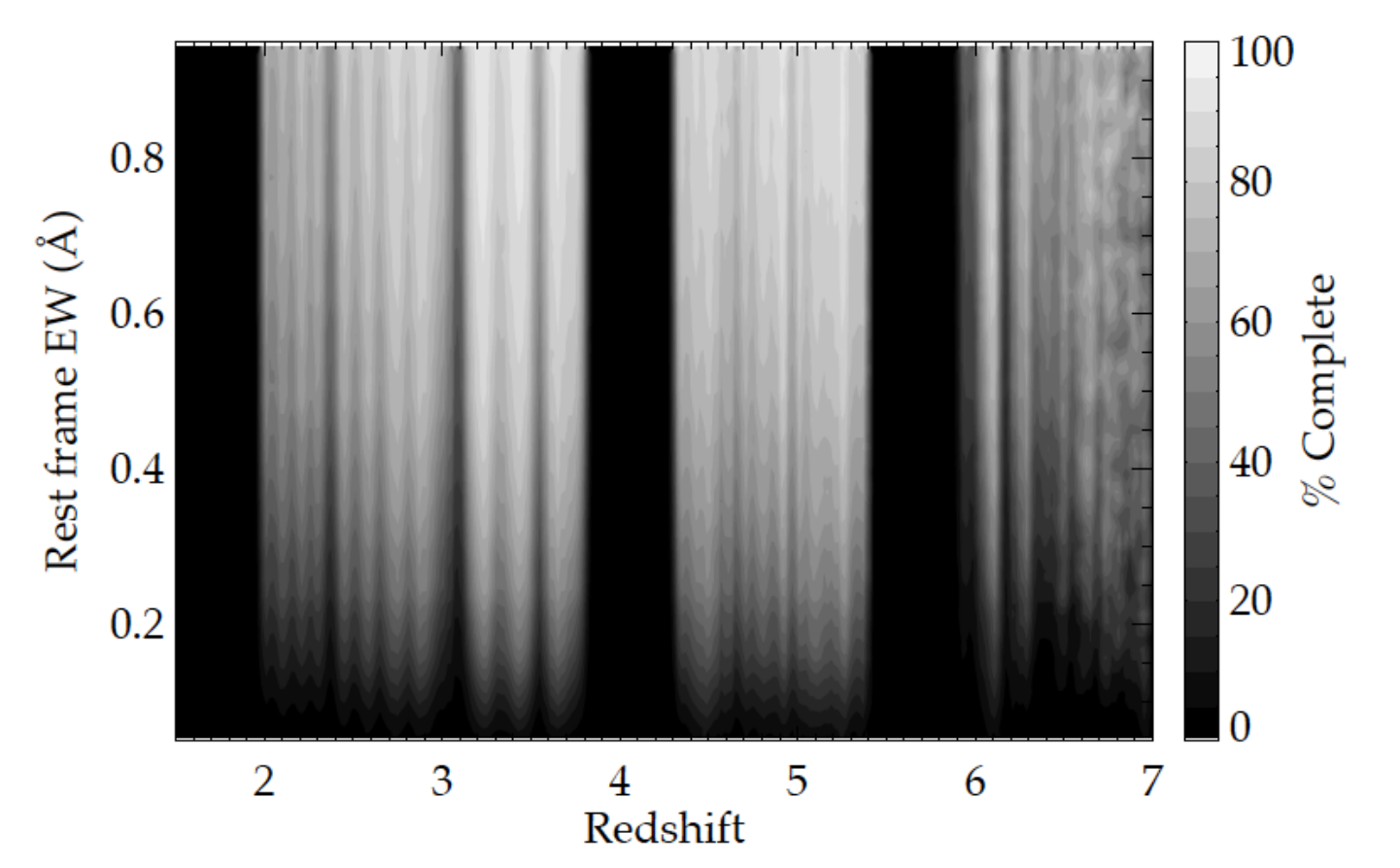}
	\end{minipage}
	\caption{The total pathlength-weighted completeness for our
          survey. Note the additional pathlength between redshifts z =
          6 and 7, which is absent from Paper I on account of its
          smaller sample size and lack of $z>6.5$ background quasars.
          The broad completeness gaps centered at $z=4.0$ and $z=5.8$
          mark the absorption bands between $J/H$, and $H/K$,
          respectively.}
	\label{fig:complete}
\end{figure}

The total pathlength-weighted completeness for our survey, thus
calculated, is shown in Figure~\ref{fig:complete}. The acceptance rate
for false positives are not included in this step, but rather are
accounted for directly in our calculations of the population
statistics.  Unlike in Paper I, the grid now extends to a
maximum redshift of $z=7$, picking up path at $z>6.1$ where \mgii
re-emerges into the $K$-band atmospheric window.

Given the completeness grid calculated above, we can calculate the
redshift path density $g(z,W)$ of our survey, i.e. the total number of
sightlines at redshift $z$ for which \mgii absorbers with equivalent
width greater than $W_r$ can be observed, as

\begin{equation}
g(z,W) = \sum_{q} R_{q}(z) C_q(z,W),
\end{equation}

where $R_q(z)$ is equal to one within the redshift search limits but
outside the redshifts excluded due to poor telluric corrections, and
zero everywhere else.  This function is shown in the top panel of
Figure 5; the bottom panel indicates the survey
path $g(W)$, defined as

\begin{equation}
g(W) = \int g(z,W) dz
\end{equation}

Here, the increase in completeness toward higher $W_r$ is reflected in
the rising path probed at larger equivalent width.  The converged
value at $g(W)\sim 150$ toward large equivalent width indicates a high
completeness, and an average redshift coverage of $\Delta z\sim 1.5$
per sightline for our 100 objects.  The total survey path of Paper I
was approximately 80, so we have roughly doubled the path by doubling
the number of QSOs observed.

\begin{figure}[t]
	\begin{minipage}{0.5\textwidth}
	\includegraphics[width=\textwidth]{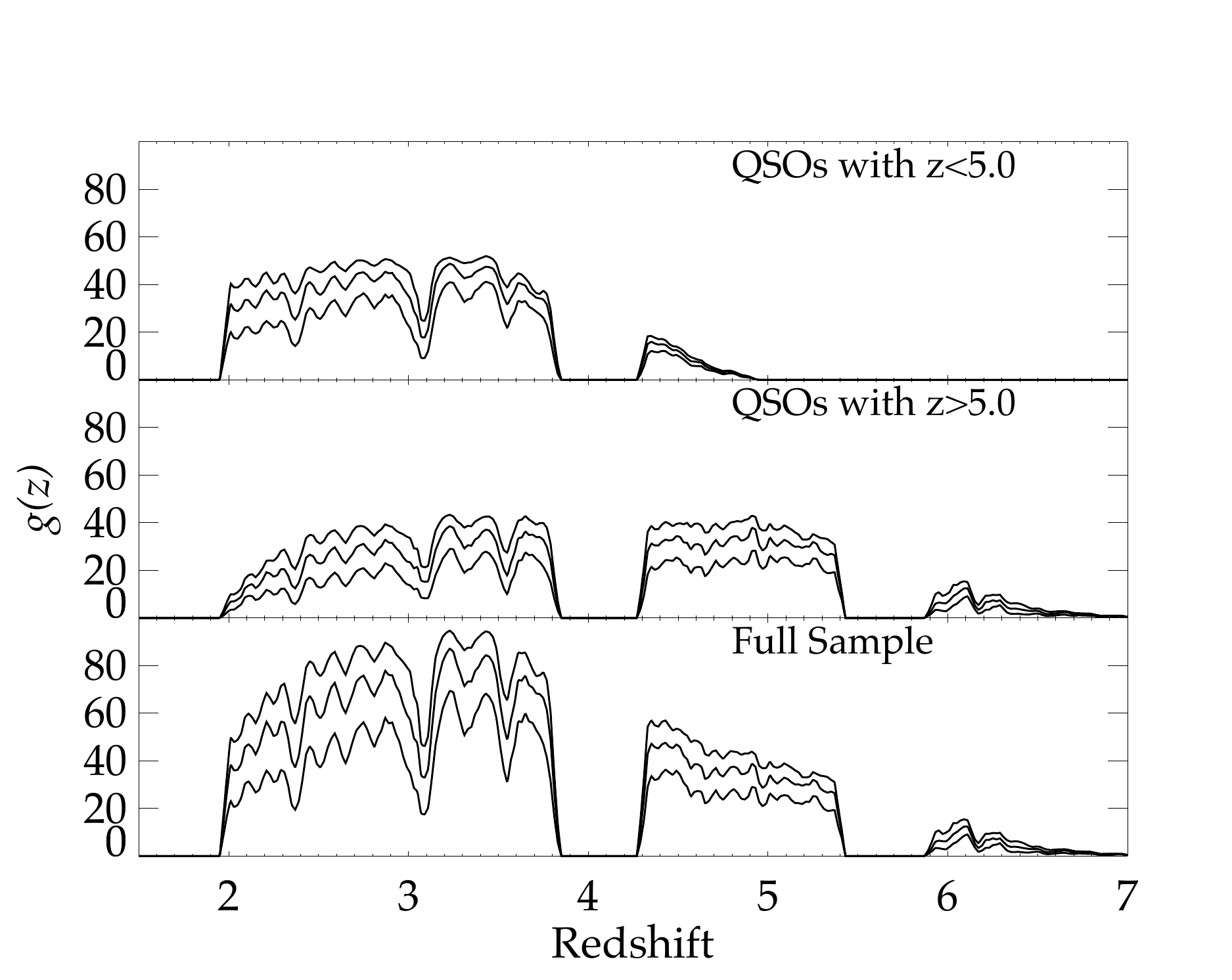}
	\includegraphics[width=\textwidth]{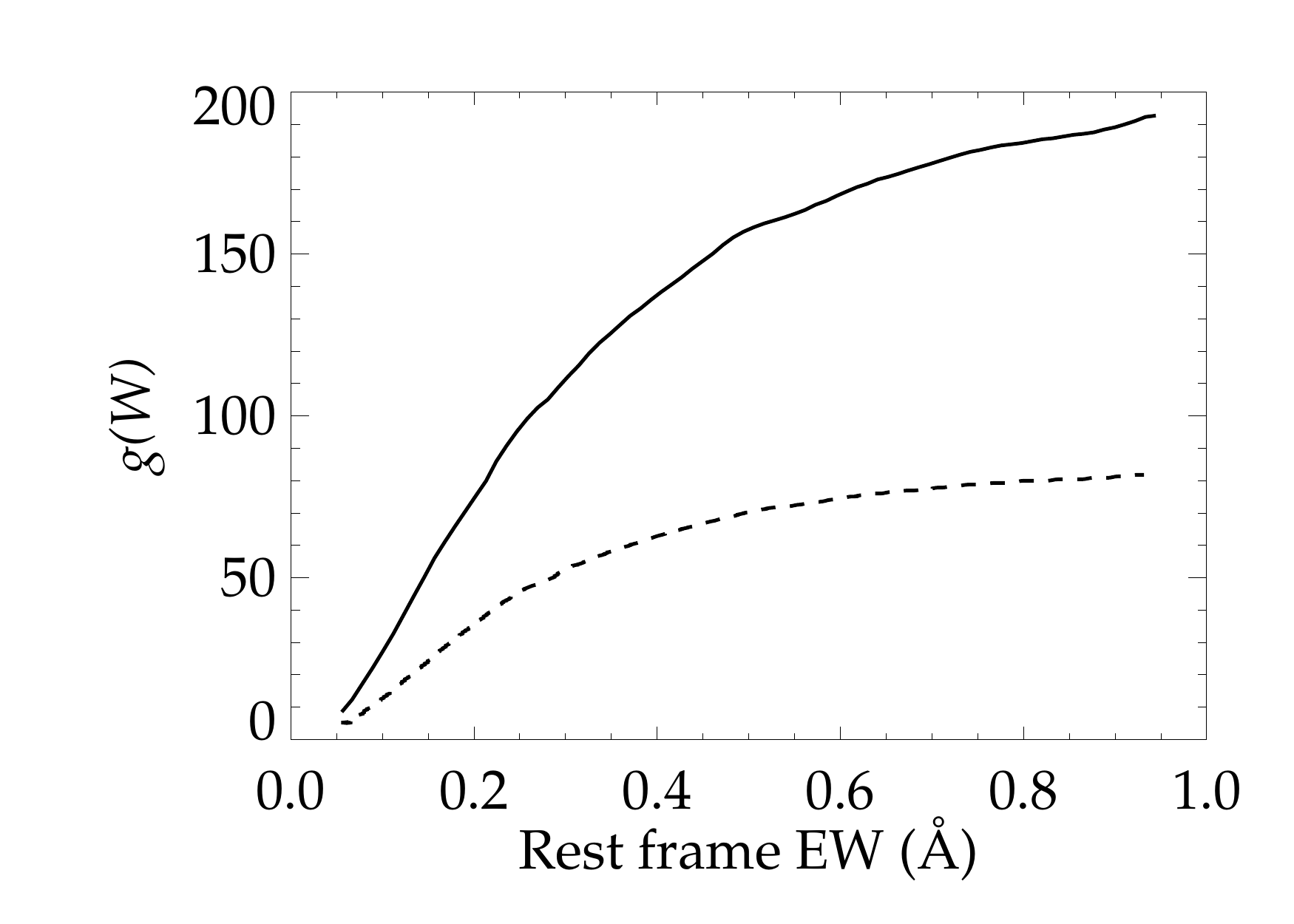}
	\caption{{\em Top}: The completeness-weighted number of
          sightlines $g(z)$ that probe the sample's redshift extent
          for three choices of $W_{R,2796}$: 0.3, 0.5, and 1.0\AA.
          {\em Bottom:} Total absorption path as a function of
          limiting $W_{r,2796}$.  These paths roughly double the
          survey volume probed by Paper I, shown as the dashed line in
          the bottom figure.}
	\end{minipage}
	\label{fig:surveypath}
\end{figure}

\section{Results}

\begin{figure}[h!]
	\begin{minipage}{0.5\textwidth}
	\includegraphics[width=\textwidth]{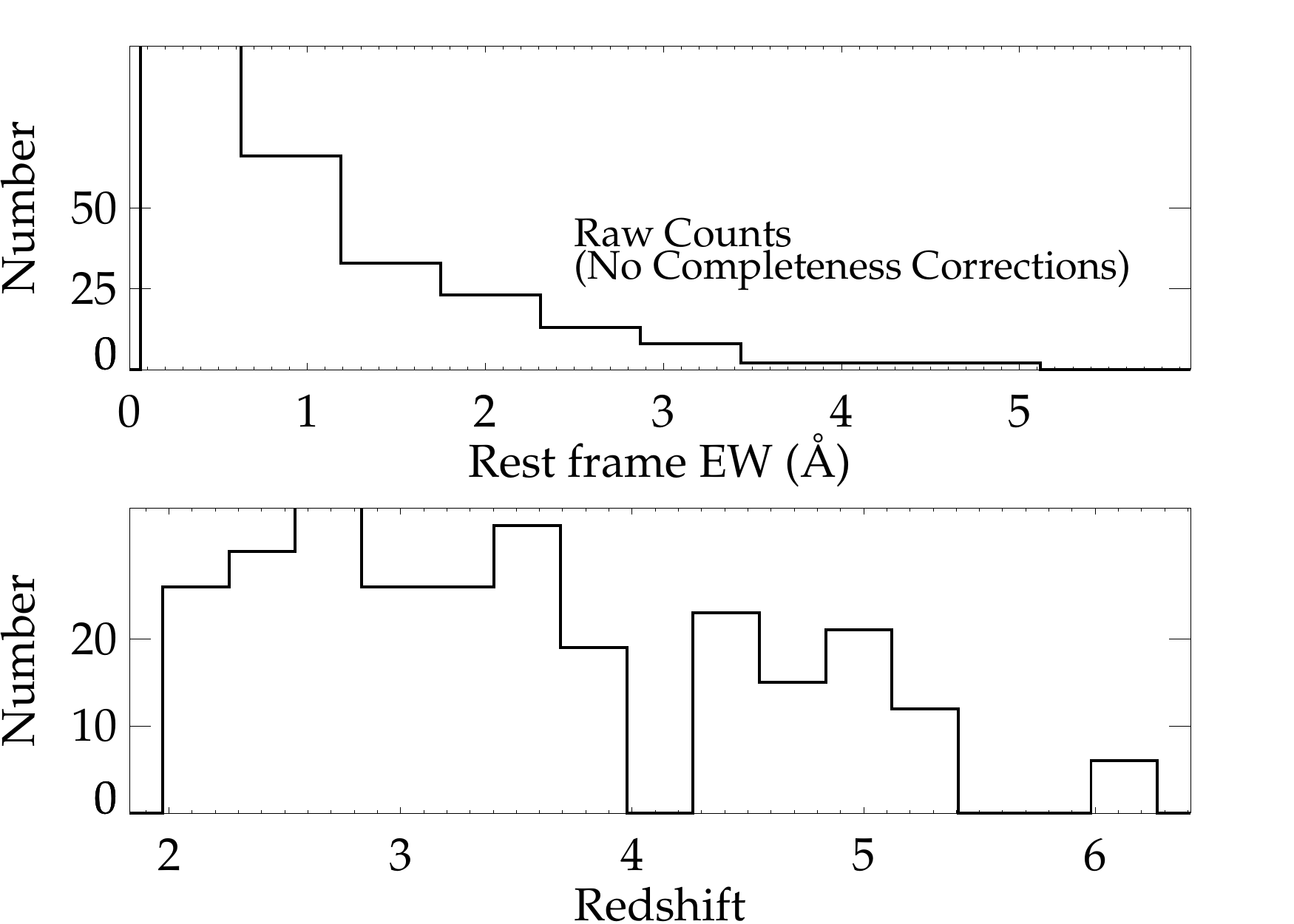}
	\end{minipage}
	\caption{Raw counts of \mgii absorbers found in our survey,
          binned by equivalent width and redshift. These numbers are
          not completeness corrected.}
	\label{fig:sample_hists}
\end{figure}

Using these methods, we identified \NumSystems ~\mgii absorbers, not
including any corrections for incompleteness. Histograms of the raw
counts of these systems based on redshift and equivalent width are
given in Figure~\ref{fig:sample_hists}.  Detailed properties of each
absorber are listed in Table \ref{tab:dlist}.

\subsection{Accounting for Completeness and False Positives}

We employed the same formalism described in Paper I, to account
for incompleteness and false positives in our statistical results.
Briefly, in a given redshift and equivalent width bin $k$, the
corrected (true) number of systems can be calculated from the number
$\breve{N}_k$ of detected systems in that bin as

\begin{equation}
N_k = \frac{\breve{N}_k(1-\bar{A}^F_k)-\bar{A}^F_k\breve{F}_k}{\bar{C}_k-\bar{L}_k \bar{A}^F_k},
\end{equation}

\noindent where $\breve{F}^k$ is the number of rejected candidates,
$\bar{C}_k$ is the average completeness, $\bar{L}_k$ is the automated
line identification finding probability, and $\bar{A}^F_k$ is the
user-acceptance rate for false positives, each calculated for the kth
bin. These fractions are calculated from the previously discussed
automated completeness tests and user-rating calibrations. An
important caveat is that the average completeness of a redshift and
equivalent width bin is weighted according to the number distribution
$d^2N/dzdW$, which must in principle be determined from the true
number of systems $N_k$. Here we follow the discussion in Paper I and
apply the simplifying assumption that $d^2N/dzdW$ is constant across
each bin to resolve the apparent circularity.

\subsection{The $W_r$ Frequency Distribution}

\begin{deluxetable}{c c c c c}
\tablecaption{\mgii Equivalent Width Distribution, \\ Full Sample and Redshift Cuts}
\tablehead{ \colhead{$\left<W_r\right>$} & \colhead{$\Delta W_r$} & \colhead{$\bar{C}$} & \colhead{Number} & \colhead{$d^2N/dzdW$} \\ \colhead{(\AA)} & \colhead{(\AA)} & \colhead{$(\%)$} & \colhead{} & \colhead{}}
\multicolumn{5}{c}{$z = 1.90-6.30$} \\
\hline
0.42 & 0.05-0.64 & 46.0 & 130 & 1.539$\pm$0.215 \\
0.94 & 0.64-1.23 & 77.7 & 66 & 0.591$\pm$0.082 \\
1.52 & 1.23-1.82 & 79.7 & 34 & 0.298$\pm$0.055 \\
2.11 & 1.82-2.41 & 79.7 & 21 & 0.185$\pm$0.042 \\
2.70 & 2.41-3.00 & 79.7 & 15 & 0.134$\pm$0.035 \\
4.39 & 3.00-5.78 & 79.7 & 14 & 0.026$\pm$0.007 \\ 
\hline

\multicolumn{5}{c}{$z = 1.95-2.98$} \\
\hline
0.42 & 0.05-0.64 & 42.0 & 56 & 1.371$\pm$0.336 \\
0.94 & 0.64-1.23 & 74.4 & 19 & 0.410$\pm$0.108 \\
1.52 & 1.23-1.82 & 76.7 & 14 & 0.308$\pm$0.089 \\
2.11 & 1.82-2.41 & 76.7 & 9 & 0.205$\pm$0.071 \\
2.70 & 2.41-3.00 & 76.7 & 8 & 0.187$\pm$0.067 \\
4.39 & 3.00-5.78 & 76.7 & 7 & 0.033$\pm$0.013 \\ 
\hline

\multicolumn{5}{c}{$z = 3.15-3.81$} \\
\hline
0.41 & 0.05-0.64 & 54.8 & 31 & 1.183$\pm$0.284 \\
0.94 & 0.64-1.23 & 85.5 & 22 & 0.691$\pm$0.152 \\
1.53 & 1.23-1.82 & 87.1 & 12 & 0.370$\pm$0.109 \\
2.12 & 1.82-2.41 & 87.1 & 6 & 0.182$\pm$0.076 \\
2.71 & 2.41-3.00 & 87.1 & 1 & 0.031$\pm$0.031 \\
4.39 & 3.00-5.78 & 87.1 & 3 & 0.020$\pm$0.011 \\ 
\hline

\multicolumn{5}{c}{$z = 4.34-5.35$} \\
\hline
0.41 & 0.05-0.64 & 52.1 & 32 & 1.840$\pm$0.397 \\
0.94 & 0.64-1.23 & 84.8 & 18 & 0.719$\pm$0.176 \\
1.53 & 1.23-1.82 & 86.6 & 6 & 0.236$\pm$0.097 \\
2.12 & 1.82-2.41 & 86.6 & 3 & 0.118$\pm$0.069 \\
2.71 & 2.41-3.00 & 86.6 & 3 & 0.118$\pm$0.069 \\
4.39 & 3.00-5.78 & 86.6 & 1 & 0.008$\pm$0.008 \\
\hline

\multicolumn{5}{c}{$z = 6.00-7.08$} \\
\hline
0.93 & 0.05-1.53 & 52.7 & 6 & 0.979$\pm$0.446 \\
2.26 & 1.53-3.00 & 68.2 & 1 & 0.131$\pm$0.132 \\
4.39 & 3.00-5.78 & 68.2 & 0 & $<$0.070
\enddata

\label{tab:dndws_var}

\end{deluxetable}

\begin{figure}[h]
	\begin{minipage}{0.5\textwidth}
	\includegraphics[width=\textwidth]{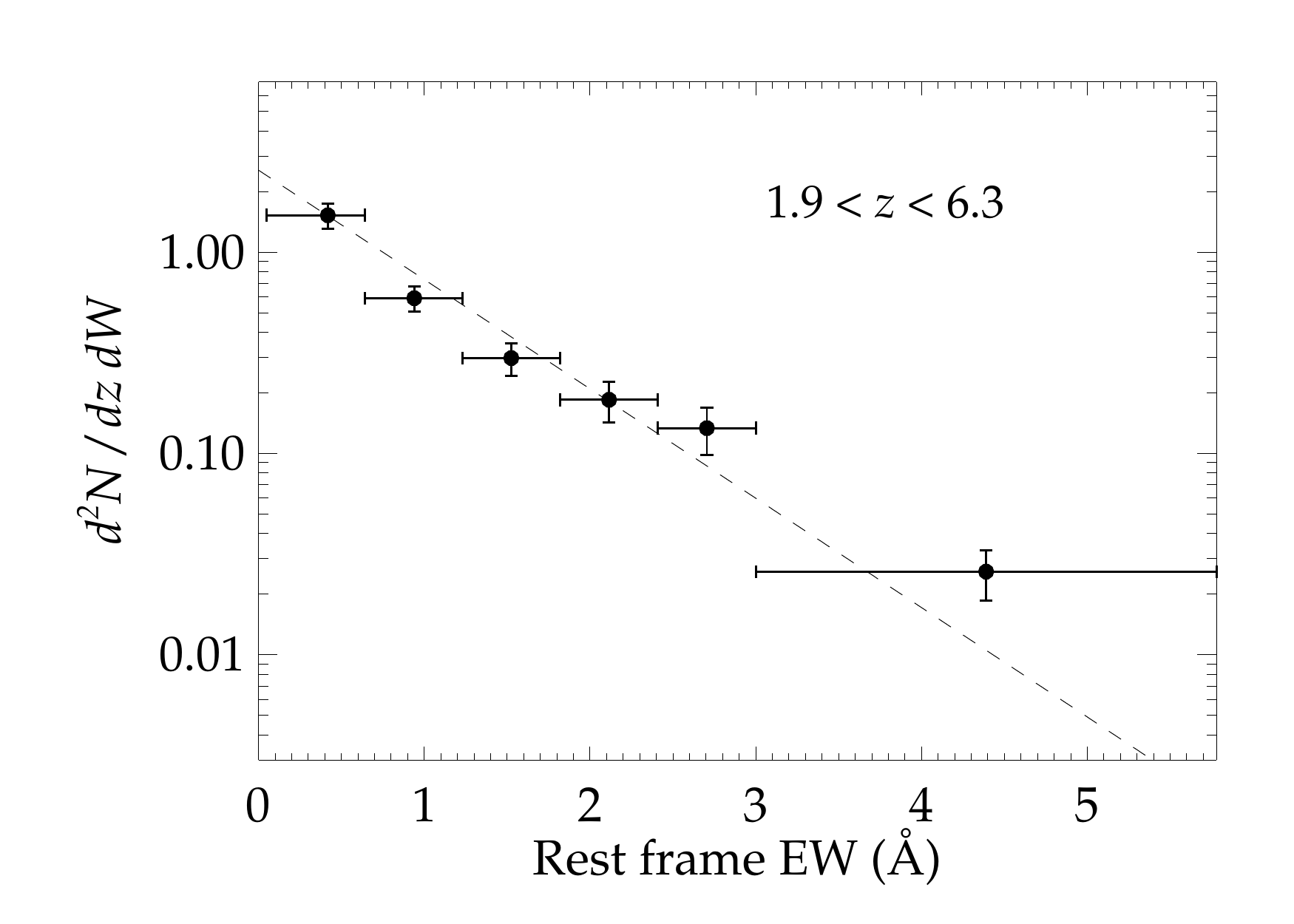}
	\caption{Population equivalent width density across the full
          survey redshift range.  The dashed line indicates a MLE fit
          for an exponential distribution.}
	\label{fig:dndws}
	\end{minipage}

\end{figure}

\begin{figure}[h]
	\begin{minipage}{0.5\textwidth}
	\includegraphics[width=\textwidth]{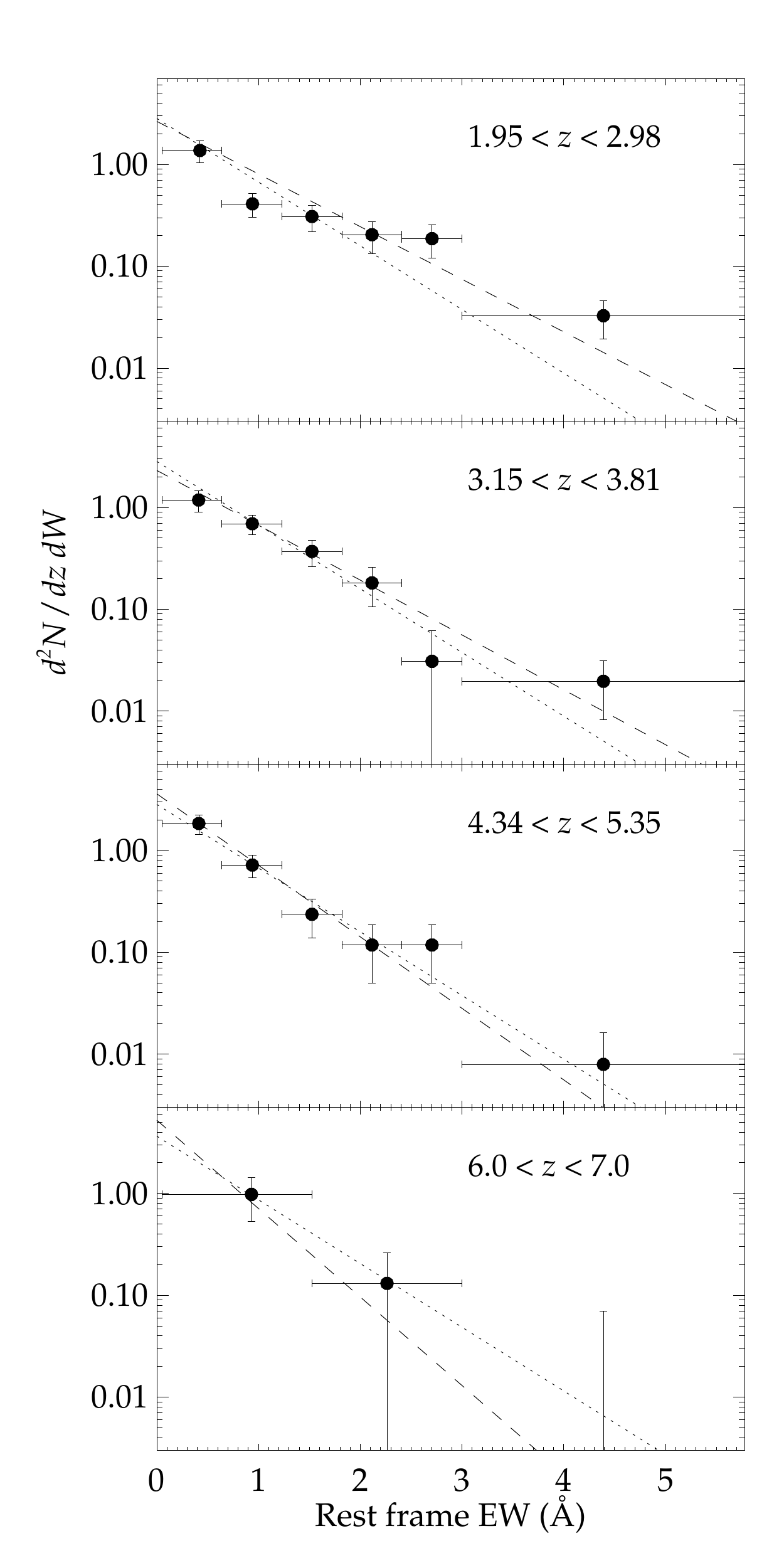}
	\end{minipage}
	\caption{Population equivalent width distribution at different
          redshift intervals. The dashed lines indicate MLE fits for
          an exponential distribution at these intervals. For comparison, the low redshift distribution from $z=1$, taken from Seyffert et al. (2013), is plotted in dotted line.}
	\label{fig:dndws_var}
\end{figure}

Table~\ref{tab:dndws_var} lists completeness-corrected values for the
rest equivalent width frequency distribution $d^2N/dzdW$, an
absorption line analog of the galaxy luminosity function. These values
are plotted in Figure \ref{fig:dndws} and Figure \ref{fig:dndws_var},
where the equivalent width distribution is binned across the full
survey redshift range and them split into four redshift intervals,
respectively.  The error bars for each point account both for Poisson fluctuations in the number count in each bin (which dominate the error budget), and for uncertainty in the completeness-adjusted path length.  The latter term reflects errors in our completeness estimates, which are much smaller by design because of the large number of simulated doublets in the completeness and user rejection tests.

With only seven systems in the highest redshift
bin, the fractional errors on each point and the associated fit
parameters are large.  However one can read off from these figures
that the density of lines at $W_r<1$\AA ~even at $z>6$ is quite
comparable to lower redshift.  At higher equivalent width ($W_r\gtrsim
2$\AA) there is weak indication of a deficit compared with lower
redshift, but the statistical errors on these points are significant
and the fit parameterizations should therefore be interpreted with
caution.

We fit the equivalent width distrbution using maximum likelihood
estimation to the exponential form

\begin{equation}
\frac{d^2N}{dzdW} = \frac{N_{\ast}}{W_{\ast}} e^{-W/W_{\ast}}
\end{equation}

\noindent by first fitting $W_{\ast}$ then setting the overall normalization
$N_{\ast}$ such that the calculated number of systems in our survey is
recovered. These fits are plotted as dashed lines in the figure of the
frequency distribution. A list of the fit parameters are given in
Table~\ref{tab:wstars}.

\begin{deluxetable}{c c c c c}
\tablecaption{Maximum-Likelihood Fit Parameters for \\ Exponential Parameterization of the $W_r$ Distribution}
\tablehead{ \colhead{$\left< z \right>$} & \colhead{$\Delta z$} & \colhead{$W^*$} & \colhead{$N^*$} \\ \colhead{} & \colhead{} & \colhead{(\AA)} & \colhead{} }
0.68\tablenotemark{a} & 0.366-0.871 & 0.585$\pm$0.024 & 1.216$\pm$0.124 \\
1.10\tablenotemark{a} & 0.871-1.311 & 0.741$\pm$0.032 & 1.171$\pm$0.083 \\
1.60\tablenotemark{a} & 1.311-2.269 & 0.804$\pm$0.034 & 1.267$\pm$0.092 \\
2.52 & 1.947-2.975 & 0.840$\pm$0.092 & 2.226$\pm$0.081 \\
3.46 & 3.150-3.805 & 0.806$\pm$0.105 & 1.864$\pm$0.069 \\
4.80 & 4.345-5.350 & 0.618$\pm$0.097 & 2.227$\pm$0.131 \\
6.29 & 5.995-7.080 & 0.500$\pm$0.148 & 2.625$\pm$0.452 \\
3.47 & 1.947-6.207 & 0.798$\pm$0.055 & 2.051$\pm$0.046
\enddata
\tablenotetext{a}{Parameter fits from \citet{nestor2005}}\label{tab:wstars}

\end{deluxetable}

\begin{figure}[t]
	\begin{minipage}{0.5\textwidth}
	\includegraphics[width=\textwidth]{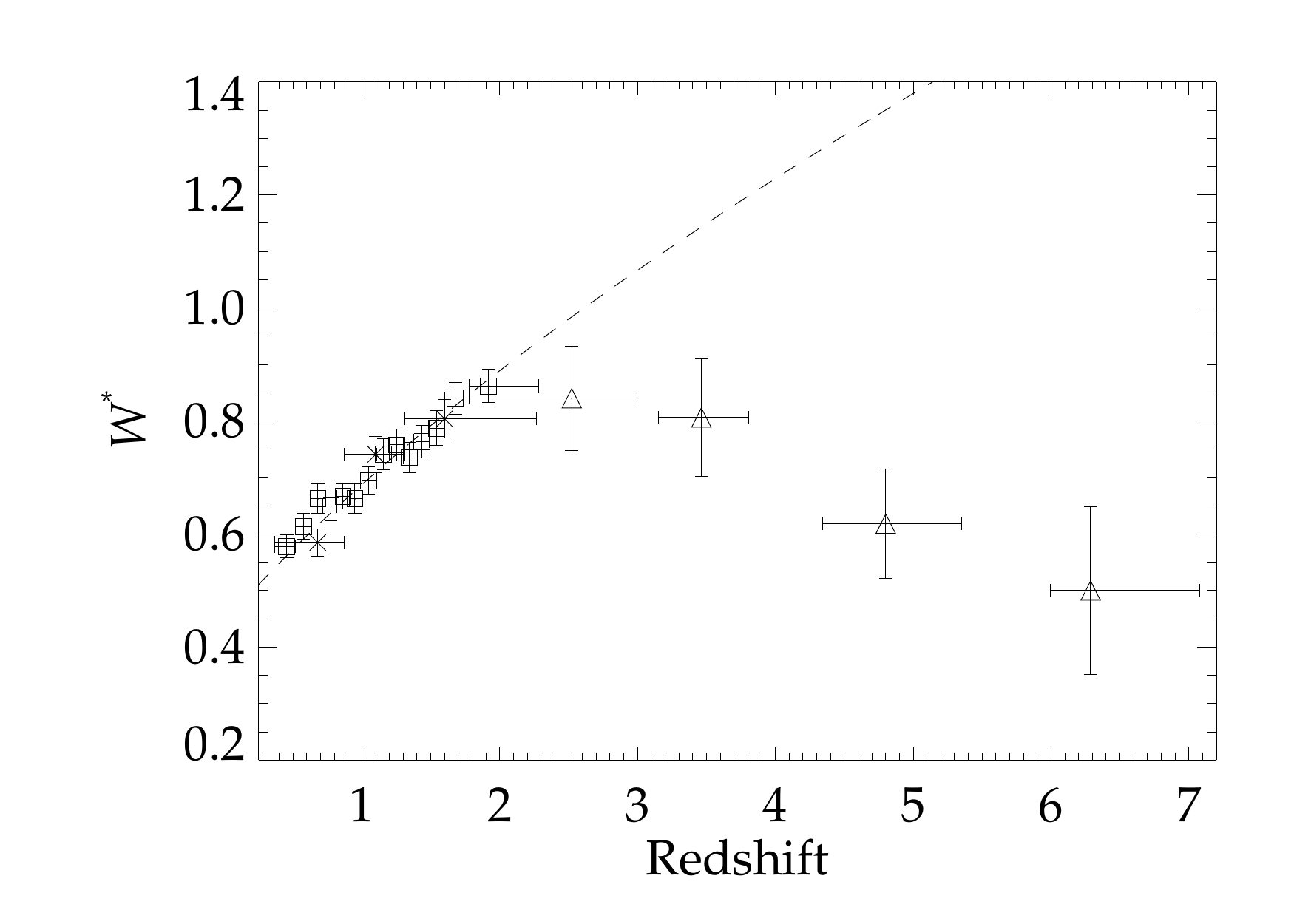}
	\end{minipage}
	\caption{The characteristic equivalent width parameter
          $W_{\ast}$ plotted by redshift. The triangles are the points
          from this survey, while the boxes and crosses show the same
          parameter for lower redshifts from \citet{seyffert} and
          \citet{nestor2005}, respectively.  The dashed line gives the
          MLE fit for this parameter from \citet{nestor2005} for the
          low redshift points.}
	\label{fig:wstar_zs}
\end{figure}

Figure~\ref{fig:wstar_zs} displays evolution in the characteristic
equivalent width $W_{\ast}$ with redshift. For comparison, we have
added the equivalent parameters provided in \citet{nestor2005} and
\citet{seyffert} at lower redshifts, though it is important to note
that Seyffert et al. only include systems with $W_r > 1$ \AA ~in their
fits. Throughout the analysis below we use these two samples as our
low-redshift refrences even though many other \mgii surveys have been
performed on the SDSS QSO sample
\citep{prochtersdss2006,lundgren2009,quiderSDSS,zhu_mgii,chen_boss_mgii,dr12_mgii}.
The main motivation for our choice is that \citet{nestor2005} probes
the smallest equivalent widths (comparable to our measurements) in the
SDSS data, while \citet{seyffert} uses identification and analysis
techniques most similar to our methods.  However these results are
broadly consistent with other works in the literature where they may
be compared.

Our results confirm the trend noted in Paper I: at higher redshifts
$W_{\ast}$ does not continue its growth with redshift at earlier
times. Rather, it peaks at around $z=2$-$3$, after which it begins to
decline.  In direct terms, this corresponds to a similar peak in the
incidence of strong \mgii absorbers around $z=2$-$3$, with a dropoff
toward early epochs in the strong systems relative to their weaker
counterparts.

\subsection{dN/dz and dN/dX}

\begin{figure}[h]
	\begin{minipage}{0.5\textwidth}%
	\includegraphics[width=\textwidth]{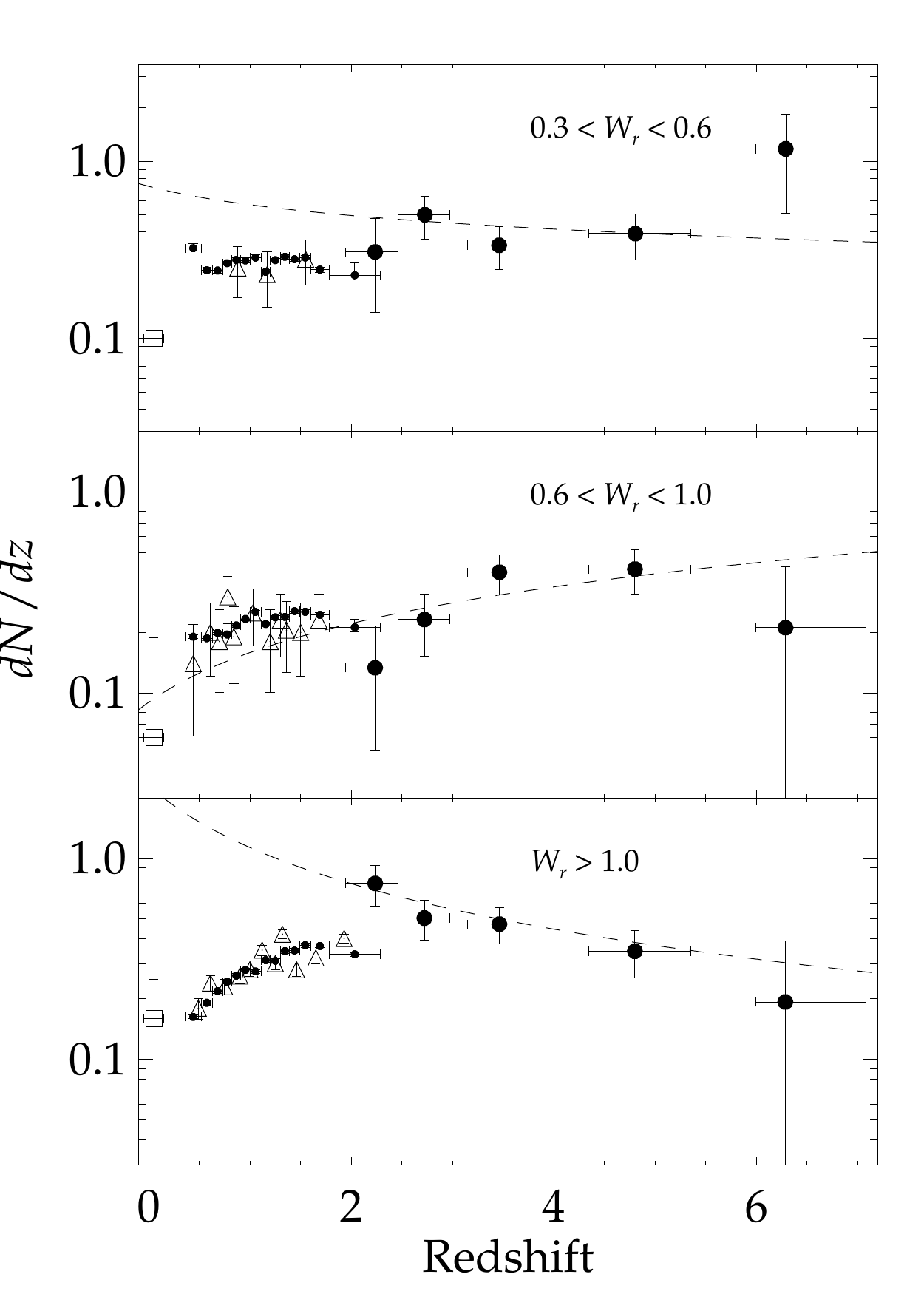}
	\caption{The line density of \mgii absorbers
          plotted by redshift, separated into three equivalent
          width ranges. Also plotted are the corresponding points at
          earlier redshift from \citet[][small black points]{seyffert}
         ,\citet[][hollow triangles]{nestor2005}, and \citet[][holllow squares]{churchill_lowz}. Dashed lines give the
          MLE fits for a power law distribution on our high redshift data.}
	\label{fig:dndzs}
	\end{minipage}
\end{figure}

\begin{figure}[h]
	\begin{minipage}{0.5\textwidth}
	\includegraphics[width=\textwidth]{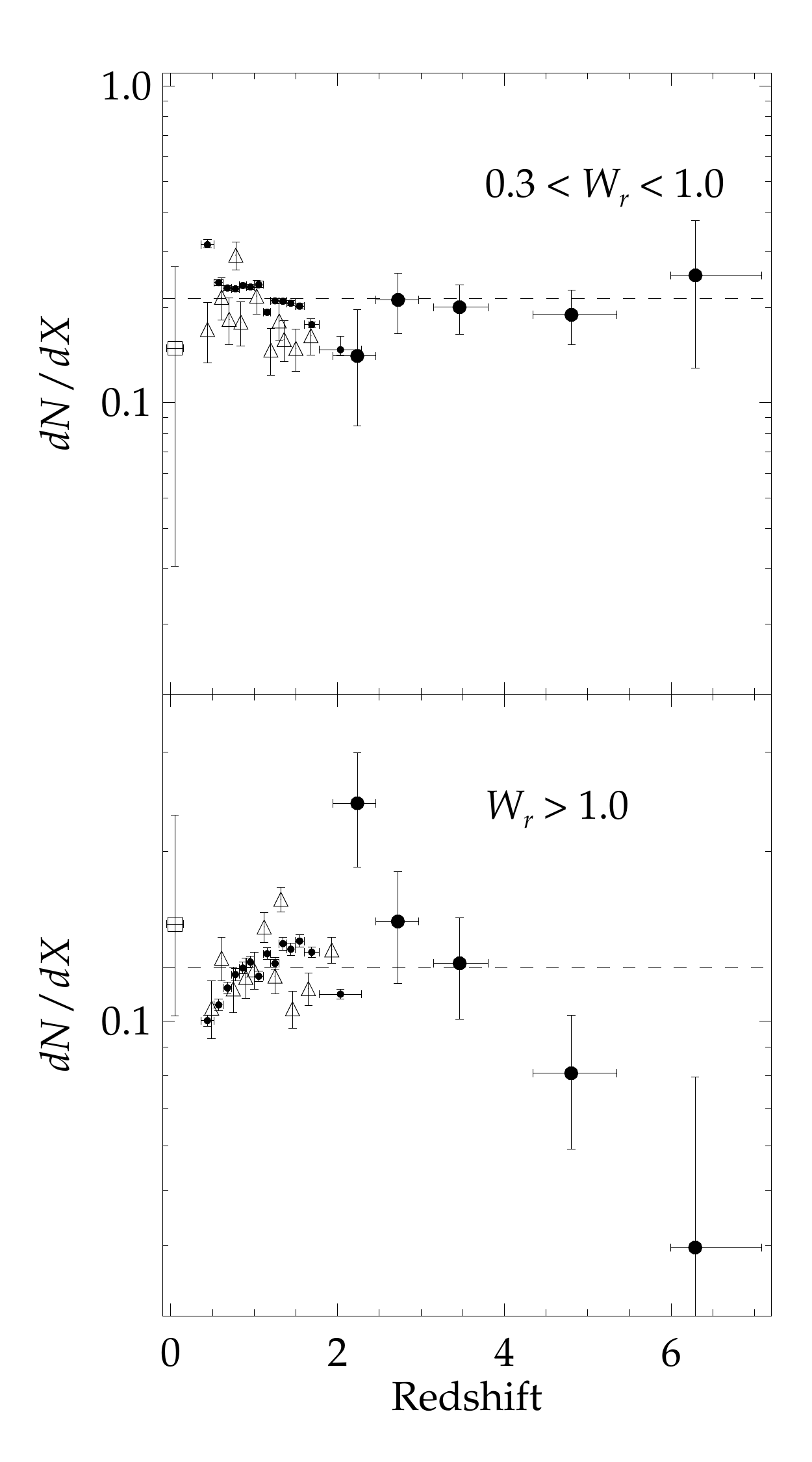}
	\caption{The comoving line density of \mgii absorbers plotted
          by redshift, separated into different equivalent width
          ranges corresponding to weak and strong absorption
          systems. Also plotted are the corresponding points at
          earlier redshift from \citet[][small black points]{seyffert}
         ,\citet[][hollow triangles]{nestor2005}, and \citet[][holllow squares]{churchill_lowz}. The dashed line
          indicates a constant comoving population.}
	\label{fig:dndxs}
	\end{minipage}
\end{figure}

The zeroth moment of the frequency distribution gives the line
density of \mgii absorption lines $dN/dz$, as plotted in Figure
\ref{fig:dndzs}. We include low-redshift points from \mgii surveys of the
SDSS \citep{nestor2005,seyffert} for comparison.
For completeness we have also performed
MLE fits of the form
\begin{equation}
\frac{dN}{dz} = N_{\ast} ( 1 + z )^{\beta}
\end{equation}
on our high redshift points, where the normalization $N_{\ast}$ is fixed such that
$dN/dz$ integrated by redshift with the survey
path density $g(z,W)$ recovers the number counts of our survey;
these fits are shown as dashed lines in Figure \ref{fig:dndzs}.
These results and parameter fits are listed in Table \ref{tab:dndzs} and
Table \ref{tab:betas}, respectively.

The line density $dN/dz$ can further be converted to the more physical
comoving line density $dN/dX$.  Here we divide the distribution into
two equivalent width bins separated at $W_r=1$\AA, and 5 redshift bins
to illustrate differences in evolutionary trends.  

As with the equivalent width distribution, error bars include a
Poisson contribution from the number of systems in each bin, and an
additional (much smaller) contribution from uncertainty in the
completeness values used to adjust the survey pathlength ($dz$ or
$dX$). The overall accuracy of the FIRE survey points is likely
limited by statistical errors---even with 100 sightlines we average
just 10-25 absorbers per bin, corresponding to a 20-30\%
uncertainty. In contrast the low redshift studies from SDSS have
thousands of absorbers per redshift bin and therefore have errors
dominated by systematic effects not explicitly quantified in these
studies (and therefore not captured in the figure). As argued by
\citet{seyffert} these likely arise from differences in (a) continuum
fitting procedures, and (b) algorithms for measuring $W_r$, and (c)
use of a sharp $W_r$ cutoff when defining samples used to derive
$dN/dX$. Comparison of different \mgii surveys from SDSS QSOs suggests
a systematic scatter of $\sim 10-15\%$, far larger than the $\sim 1\%$
random errors \citep{seyffert} These different errors must be
considered when comparing in regions of overlap such as the bottom
panel of Figure 11.

The larger survey confirms and strengthens two key findings of
Paper I by both reducing Poisson errors on points at $z<5$,
and adding new redshift coverage at $z>6$.  First, the comoving
absorption density (i.e. the frequency) of typical \mgii systems with
$0.3<W_r<1$\AA ~remains remarkably constant from $z=0.5$ to $z=7$,
i.e. all redshifts that have been searched.  

This can only be true if the product of the comoving volume density of
absorbers $n(z)$, multiplied by the physical cross section of each
absorber $\sigma$, also remains a constant.  If \mgii absorbers at
high redshift are associated with luminous galaxies like their
low-redshift counterparts, then circum-galactic gas must therefore
have a substantial cross-section for heavy-element absorption even
very early in these galaxies' evolutionary history.  Our previous work
suggested this result to $z=5.5$; the new sightlines presented here
exhibit the exact number of \mgii one would expect from simple
extrapolation of this trend to $z=6.5$, when the universe was $850$
Myr old.

The second key finding from Paper I confirmed here is a firm
evolution in the frequency of strong \mgii absorbers at $W_r>1$\AA.
This trend is in marked contrast to the weaker systems, and is
consistent with the evolution in $W^*$ of the frequency distribution
$d^2N/dXdW$.  We find just one strong system at $z>6$, again
consistent with expectations extrapolated from lower $z$.  The decline
of nearly an order of magnitude from the peak at $z\sim 2.5$ suggests
that further searches for strong systems at $z\sim 7$ and beyond are
likely to require many sightlines toward faint QSOs; however the
weaker systems may well remain plentiful.

\begin{deluxetable}{c c c c c c}
\tablecaption{\mgii Absorption line density $dN/dz$}
\tablehead{ \colhead{$\left< z \right>$} & \colhead{$\Delta z$} & \colhead{$\bar{C}$} & \colhead{Number} & \colhead{$dN/dz$} & \colhead{$dN/dX$} \\ \colhead{} & \colhead{} & \colhead{$(\%)$} & \colhead{} & \colhead{} & \colhead{} }
\startdata

\multicolumn{6}{c}{$0.3 \angmath <W_r<0.6$ \angmath}\\
\hline
2.236 & 1.947-2.461 & 46.2 & 11 & 0.308$\pm$0.168 & 0.100$\pm$0.054 \\
2.727 & 2.461-2.975 & 62.2 & 18 & 0.499$\pm$0.137 & 0.149$\pm$0.041 \\
3.460 & 3.150-3.805 & 69.4 & 15 & 0.336$\pm$0.091 & 0.090$\pm$0.024 \\
4.806 & 4.345-5.350 & 66.2 & 13 & 0.391$\pm$0.113 & 0.091$\pm$0.026 \\
6.291 & 5.995-7.085 & 43.1 & 4 & 1.173$\pm$0.667 & 0.241$\pm$0.137 \\ 
\hline
\multicolumn{6}{c}{$0.6 \angmath<W_r<1.0 \angmath$}\\
\hline
2.236 & 1.947-2.461 & 63.7 & 5 & 0.133$\pm$0.081 & 0.043$\pm$0.026 \\
2.723 & 2.461-2.975 & 78.9 & 10 & 0.232$\pm$0.079 & 0.069$\pm$0.024 \\
3.463 & 3.150-3.805 & 83.7 & 21 & 0.398$\pm$0.090 & 0.107$\pm$0.024 \\
4.802 & 4.345-5.350 & 82.6 & 17 & 0.412$\pm$0.103 & 0.096$\pm$0.024 \\
6.289 & 5.995-7.085 & 62.5 & 1 & 0.211$\pm$0.213 & 0.043$\pm$0.044 \\ 
\hline
\multicolumn{6}{c}{$W_r>1.0 \angmath$}\\
\hline
2.236 & 1.947-2.461 & 68.7 & 24 & 0.751$\pm$0.172 & 0.244$\pm$0.056 \\
2.722 & 2.461-2.975 & 83.4 & 22 & 0.505$\pm$0.113 & 0.150$\pm$0.034 \\
3.463 & 3.150-3.805 & 87.1 & 26 & 0.471$\pm$0.096 & 0.127$\pm$0.026 \\
4.801 & 4.345-5.350 & 86.6 & 15 & 0.346$\pm$0.092 & 0.081$\pm$0.022 \\
6.287 & 5.995-7.085 & 68.2 & 1 & 0.193$\pm$0.195 & 0.040$\pm$0.040
\enddata

\label{tab:dndzs}

\end{deluxetable}

\begin{deluxetable}{c c c c c}
\tablecaption{Maximum-Likelihood Estimates of the \\ Line Density Evolution $dN/dz=N^*(1+z)^\beta$}
\tablehead{ \colhead{$\left< W_r \right>$} & \colhead{$\Delta W_r$} & \colhead{$\Delta z$} & \colhead{$\beta$} & \colhead{$N^*$} \\ \colhead{(\AA)} & \colhead{(\AA)} & \colhead{} & \colhead{} & \colhead{} }
\startdata
1.17\tablenotemark{a} & 1.00-1.40 & 0.35-2.3 & $0.99^{+0.29}_{-0.22}$ & $0.51^{+0.09}_{-0.10}$ \\
1.58\tablenotemark{a} & 1.40-1.80 & 0.35-2.3 & $1.56^{+0.33}_{-0.31}$ & $0.020^{+0.05}_{-0.05}$ \\
1.63\tablenotemark{a} & 1.00+ & 0.35-2.3 & $1.40^{+0.16}_{-0.16}$ & $0.08^{+0.15}_{-0.05}$ \\
2.08\tablenotemark{a} & 1.40+ & 0.35-2.3 & $1.74^{+0.22}_{-0.22}$ & $0.036^{+0.06}_{-0.06}$ \\
2.52\tablenotemark{a} & 1.80+ & 0.35-2.3 & $1.92^{+0.30}_{-0.32}$ & $0.016^{+0.06}_{-0.03}$ \\
0.45 & 0.30-0.60 & 1.9-6.3 & -0.345$\pm$0.616 & 0.722$\pm$0.653 \\
0.79 & 0.60-1.00 & 1.9-6.3 & 0.821$\pm$0.505 & 0.090$\pm$0.069 \\
1.80 & 1.00+ & 1.9-6.3 & -1.020$\pm$0.475 & 2.298$\pm$1.561
\enddata\tablenotetext{a}{Parameter fits from \citet{prochtersdss2006}, with corresponding upper and lower $95\%$ confidence intervals.  This survey's results include $1\sigma$ errors.}\label{tab:betas}

\end{deluxetable}

\subsection{Comparison with Other Searches for High-Redshift MgII}
In the time since initial submission of this paper, two other relevant manuscripts have been posted describing \mgii searches in the near-IR. We comment briefly here on comparisons of these studies with our work.

\citet{codoreanu2017} searched a sample of four high-SNR spectra
obtained with VLT/XShooter for \mgiinsp; because of the exceptional
data quality this search is more sensitive to weak absorption lines
but its shorter survey path length leads to larger Poisson
uncertainties in bins of higher equivalent width.  In the regions
where our samples are best compared ($0.3<W_r<1.0$\AA) the agreement
in number density is very good. Our larger sample size reveals
evidence for evolution at $W_R>1$\AA ~not visible in their data;
however, their higher sensitivity reveals numerous weak systems
($W_r<0.3$\AA).  While we report some such systems in Table 2, our
overall completeness was not sufficient to claim robust statistics on
these absorbers. Their analysis reveals an excess of weak \mgii
systems relative to an extrapolation of the exponential frequency
distribution, as found at lower redshift.  The trend of number density
with redshift for these weak systems is broadly consistent with no
evolution, though increased sample size could reveal underlying
trends.

Separately, \citet{bosman2017} performed an ultra-deep survey for
\mgii along the line of sight to ULAS1120+0641, also covered in our
sample. They recover the two systems in our sample, and further
recover three systems with $W_r<0.3$\AA ~at $z>6$ not detected by our
search (because of our lower SNR, particularly in regions of strong
and/or blended telluric absorption and emission).  The number of weak
systems uncovered in this sightline tentatively suggests that the
frequency distribution may transition to a power-law slope at low
column densities where our survey would have correspondingly low
completeness.

\section{Discussion}

We have extended the original survey of Paper I from 46 to \NumQSOs
~QSOs, with particular emphasis on increasing path length at higher
redshift. While significantly augmenting the sample of \mgii
absorbers, we confirm the trends noted in Paper I. Our data (1) rule
out the monotonic growth of $W_{\ast}$ at high redshifts and (2) show
that the comoving line density of $W_r<1$\AA ~\mgii absorbers 
does not evolve within errors, while
stronger absorbers demonstrate a noticeable decline in comoving line density. In particular, our
detection of five \mgii systems at $z >6$ with equivalent width
$0.3<W_r<1.0$\AA ~conforms with a constant comoving population
ansatz for the weak \mgii systems.

\subsection{Strong \mgii and the Global Star Formation Rate}

In Paper I, we discussed the hypothesis that strong \mgii absorption
is linked closely with star forming galaxies, using the scaling
relation presented in \citet{menardo2} to convert \mgii equivalent
widths into an effective contribution to the global star formation
rate.  This integral is dominated by the strongest absorbers in the
sample, which peak strongly in number density near $z\sim 2$-$3$,
similar to the SFR rate density.

The conversion method relies on a correlation observed in SDSS-detected \mgii systems between $W_r$ and \oii luminosity surface density measured in the same fiber as the background QSO: 
\begin{equation}
\left<{\Sigma_{L_{OII}}}\right> \propto \left( {{W_r}\over{1{\rm\AA}}}\right)^{1.75}.
\end{equation}
By integrating the \mgii equivalent width distribution $d^2N/dWdX$, weighted by the function in Equation 9, one obtains a volumetric luminosity density of \oii which can then be converted into a star formation rate density using the \oii~- SFR scaling relations of \citet{zhuSFR}. As discussed in Paper I, one should keep in mind the possibility raised by \citet{lopezChen} that the correlation in Equation 9 could arise from the decline in $W_r$ with impact parameter, coupled with differential loss of \oii flux from the SDSS fiber, rather than a physical link between the SFR and \mgii absorption strength. However in light of the observed evolution in $dN/dX$ for strong systems, the large velocity spreads seen in the strongest systems, and the link between star formation and \mgii seen in individual galaxies \citep{bouche2007ha,noterdaeme2010}, we explore this possibility while acknowledging its possible limitations.

\begin{figure}[h]
	\begin{minipage}{0.5\textwidth}
	\includegraphics[width=\textwidth]{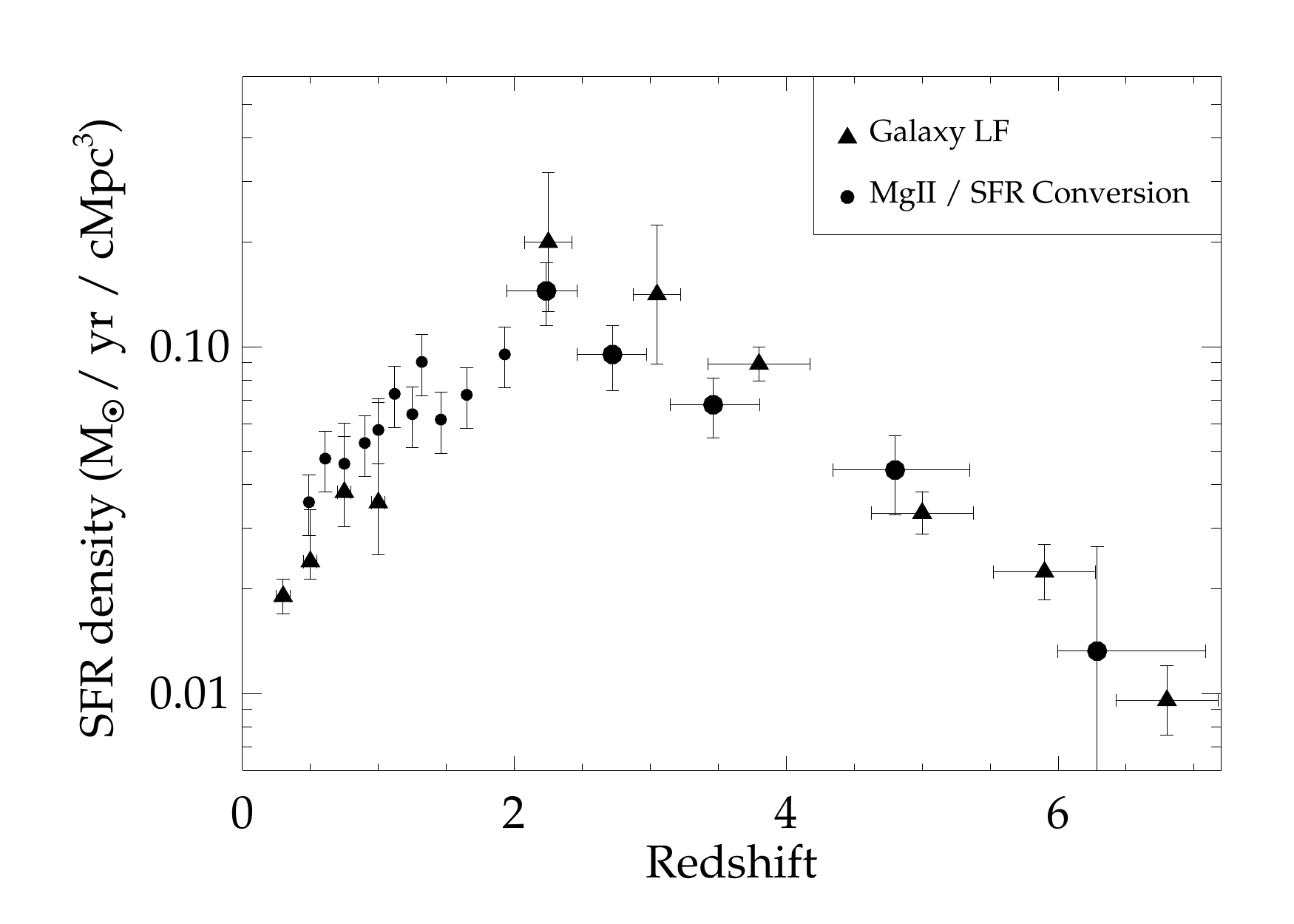}
	\caption{Comparison of the SFR density in units of comoving
          Mpc$^3$, determined directly from observations of deep
          fields (triangles) and as converted from \mgii absorber
          statistics using the prescription of \citet{menardo2}.  The
          coincidence remains in place with an increased sample size
          and the addition of a new point at higher redshift.}
	\label{fig:sfr}
	\end{minipage}
\end{figure}

Figure~\ref{fig:sfr} presents an updated version of this calculation,
with smaller errors from our new and larger sample, and an additional
point at $z>6$ from our new high redshift sightlines.  Despite the
caveats presented in Paper I about the methodology of the \mgii-SFR
conversion \citep{lopezChen}, and the application of low-redshift
scalings at these early epochs, the agreement between the
\mgiinsp-inferred SFR and the values measured directly from deep fields
remains remarkable. This suggests that at least the strongest \mgii
systems in our surveys derive their large equivalent widths
(i.e. their velocity structure) from processes connected to star
formation.

\subsection{Low Mass Halos as Sites of Early \mgii Absorption}

The persistence of \mgii at $dN/dX\approx 0.2$ absorbers per
comoving path length at $z\sim 5.5$-$6.5$ merits further examination,
because it implies that some CGM gas was enriched very early in cosmic
history---indeed, well before galactic stellar populations were fully
relaxed.  In Paper I we explored whether known high-redshift galaxy
populations could plausibly account for the observed number of \mgii
systems, supposing that radial scaling relations of $W_r$ and covering
fraction measured at $z<0.5$ apply at early times.  These calculations
essentially integrate down a mass function or a luminosity function to
obtain a number density of halos, and then seed these with \mgii gas
using a radial prescription.  \mgii absorption statistics calculated
in this way are sensitive to the lower limit of integration, as well
as the value assumed for the low-mass (or faint-end) slope.

In that work, we first examined the predictions of a halo-occupation
distribution model from \citet{chenandtinker2009}.  These authors integrate
the halo mass function down to a fixed, redshift-independent cutoff
below which it is assumed that galaxies do not harbor \mgii in their
CGM.  The cutoff is chosen to match the evolution in number statistics
below $z<2$, but substantially underpredicts the \mgii incidence rate
at higher redshift.  This likely results from the evolving mass
function; since halos have lower masses at early times, a higher
percentage of galaxies miss the (redshift-independent) mass cut.

\begin{figure}[h]
	\begin{minipage}{0.5\textwidth}
	\includegraphics[width=\textwidth]{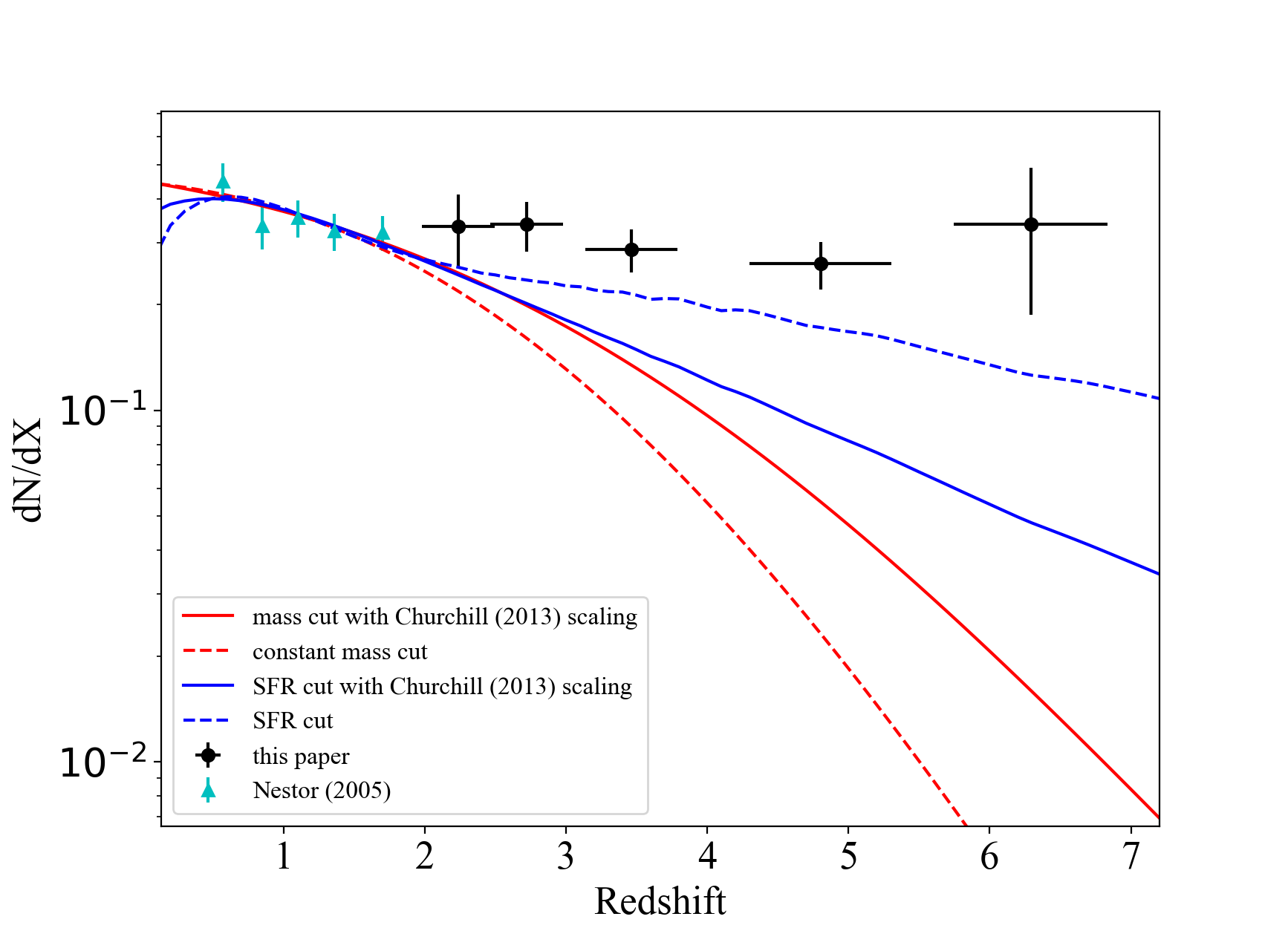}
	\caption{Comparison of the comoving absorber number density with $W_r > 0.3 \AA$
          with halo statistics from Illustris.  The green triangles at
          low redshift are from \citet{nestor2005}, black points are
          from this paper. The red and blue curves correspond to integrating the dark matter halo mass function down to a fixed, redshift-independent mass or to a fixed star formation rate, respectively. The solid and dotted lines respectively denote models in which halos have 50$\%$ covering fraction out to $R=90$ proper kpc or total cross sections that scale with dark matter mass as in 
         \citet{churchill_rvir}. The mass functions are taken from \citet{torrey_model}. The accounting of SFR vs halo mass are
          based on \citet{behroozi}}
	\label{fig:sfr_massfuncs}
	\end{minipage}
\end{figure}

In the red lines of Figure \ref{fig:sfr_massfuncs} we approximate this
calculation, using the dark matter mass function extracted from
the Illustris cosmological simulation \citep{torrey_massfunc}. We consider two 
models for halo cross section.  The first, simplest model specifies that all halos have 
a constant absorption radius of $90$ proper kpc, and geometric covering factor $\kappa=0.5$ within this volume (dashed red line). The second model (solid red line) assumes that a halo's absorption radius scales with mass as in \citet{churchill_rvir}.  We integrate the cross-section weighted mass function for each model down to a redshift-independent mass cut that matches the low redshift line densities of \citet{nestor2005}. These two scalings require mass cuts at $10^{11}$ and $10^{10}$ solar masses, respectively. This
model is only slightly simpler than that of \citet{tinker_chen}, who also 
included a radial scaling of $W_r$ and a varying absorption efficiency with halo mass.  However we verified that both methods reproduce the same basic result: strict allocation of \mgii absorption by halo mass underpredicts $dN/dX$ at
high redshift.  

\begin{figure}[h]
	\begin{minipage}{0.5\textwidth}
	\includegraphics[width=\textwidth]{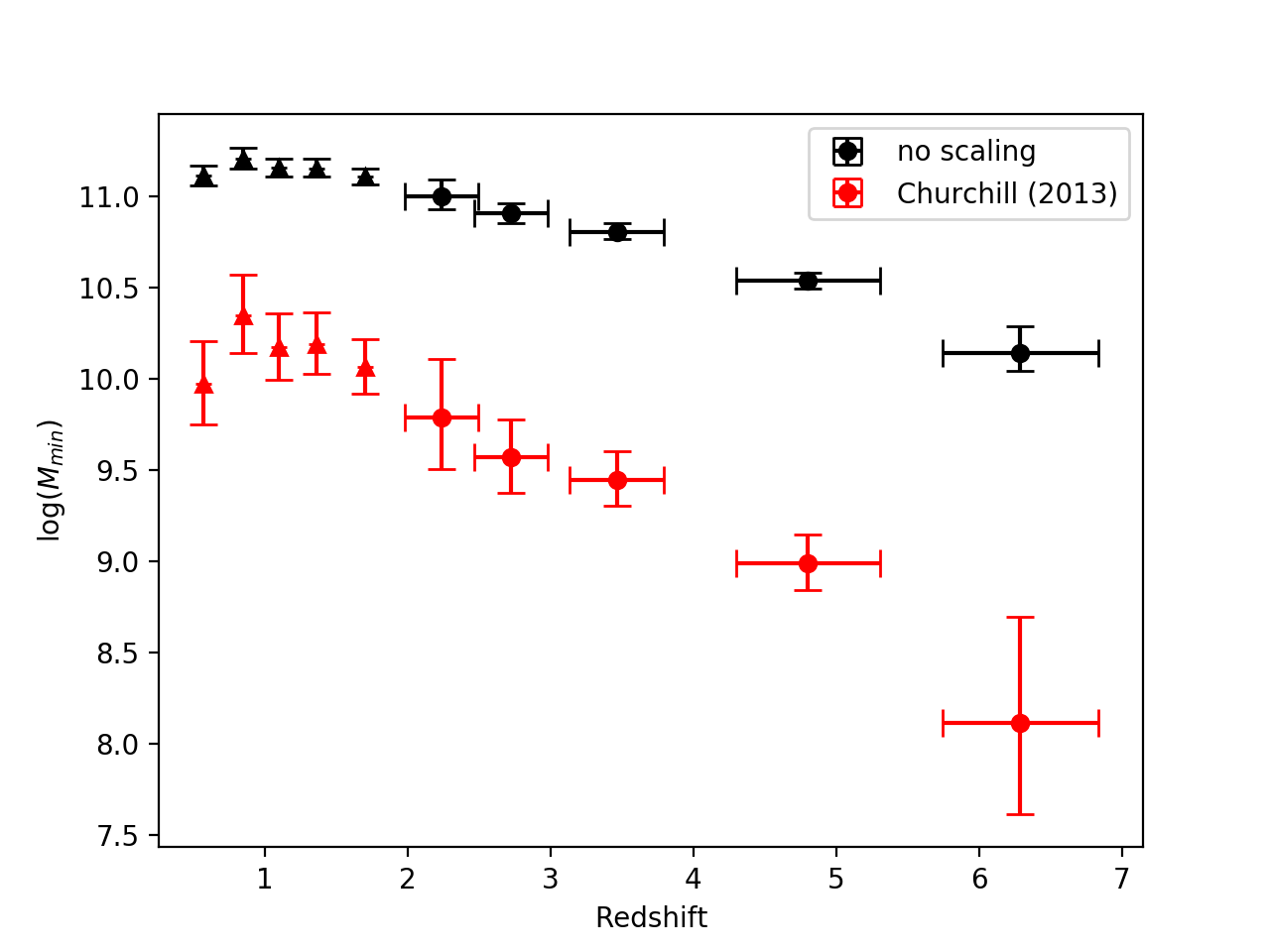}
	\caption{Minimum dark matter halo mass cuts required to reproduce the observed comoving absorber line densities, which decline by an order of magnitude or more in our survey's redshift range. The black and red points give the mass cut in models where halos have a fixed, $90$ kpc cross section with 50$\%$ covering fraction and where halo cross sections scale as in \citet{churchill_rvir}, respectively. The triangular points at lower redshifts correspond to data from \citet{nestor2005}, and the circular points derive from our data.}
	\label{fig:mmins}
	\end{minipage}
\end{figure}

Alternatively, one can specify a parameterized halo geometry and then explore how far down one must set the minimum mass limit of integration to reproduce the flat trend of $dN/dX$ for that model.  The evolution of this minimum integration mass is shown in Figure \ref{fig:mmins}, again for a fixed halo radius of $R=90$ proper kpc (red points) and using Churchill's mass-radius scaling (black points). The minimum required halo mass declines by two orders of magnitude between redshifts $z=1-6$ when cross sections scale as in \citet{churchill_rvir}, and by an order of magnitude even when cross sections do not scale with mass.  If these radial scalings apply at early times, then the observed incidence rate of \mgii requires absorption from smaller mass halos in the early universe.

While straight halo mass cuts are conceptually simple, we are not
limited to this criterion. Numerous authors have investigated 
the density of halos as a
function of both stellar mass and star formation rate (SFR).  In fact for
Paper I we found that $dN/dX$ was reproduced better at high redshift
using weighted integrals of the luminosity function rather than the
mass function.  This was a purely empirical calculation, which used
observed luminosity functions that required corrections for
observations different redshifts, filters, and systematic survey
completeness.

In Illustris, we have additional direct access to the star formation
history of each simulated galaxy.  Since the average SFR at fixed
halo mass is larger at earlier times \citep{behroozi}, we may
integrate instead down to a fixed, redshift-independent SFR, which
corresponds to lower dark matter halo mass at higher redshift.  This achieves the desired effect of seeding smaller halos with \mgii at early times.  

The blue lines in Figure \ref{fig:sfr_massfuncs} show the result of
this calculation for Illustris, using a constant-radius $90$ pkpc halo for objects above SFR$>0.5M_\odot$/yr (blue dashed line) and Churchill's mass-dependent radial scaling for objects with SFR$>0.02M_\odot$/yr (solid blue line). As before the minimum SFRs are selected to fit low redshift \citep{nestor2005} measurements.  This methodology
increases $dN/dX$ by an order of magnitude or more at high redshifts,
partially mitigating the discrepancy with a redshift-independent, fixed-mass bound on the integration. However there is no single value for the SFR cutoff that fits all redshifts; the value chosen here is a compromise but predicts too many \mgii absorbers at low redshift and slightly too few at early times.  

\subsubsection{Can Low-Mass Galaxies Yield Enough Magnesium to Enrich Gaseous Halos?}

At face value the small halo masses at high redshift in Figure 14---corresponding to even
smaller stellar masses---require us to consider whether the these
objects' stellar populations could plausibly produce enough magnesium
to fill their intra-halo media at the radii required for
observation.

If we define the {\em galaxy} yield $\eta$ as the ratio of magnesium
mass in the circumgalactic halo to the galaxy's stellar mass, then for
a mass-independent halo radius:

\begin{eqnarray}
\eta &\equiv& {{M_{Mg}}\over{M_*}} \\
&=& \frac{1}{M_*} \kappa \pi R^2 N_{\mmgii} f^{-1} m_{\mmgii} \nonumber\\
&=& 2.5 \times 10^{-5} \left(\frac{R}{90
\text{ kpc}}\right)^2 \left(\frac{M_*}{10^{10} M_{\odot}}\right)^{-1}
\left(\frac{N_{\mmgii}}{10^{13} \text{ cm}^{-2}}\right) \nonumber \\
& & \times
\left(\frac{\kappa}{0.5}\right) \left(\frac{f}{0.1}\right)^{-1},\nonumber
\end{eqnarray} 

where $\kappa$ represents the \mgii absorption covering factor, $R$ is
the gaseous halo radius, $f$ is the \mgii ionization fraction,
$m_{\mmgii}=24.3u$ is the mass of a magnesium ion, and
$N_{\mmgii}=10^{13}$cm$^{-2}$ represents the typical column density of a
modestly saturated absorption component. For our lowest-mass halos at
$z\sim 6-7$, Figure 13 provides a lower integration limit of $M_h\sim
10^{10.2}M_{\odot}$, corresponding to a stellar mass of $10^8 M_\odot$
\citep{behroozi}.  For these inputs the required {\em galactic} yield of
$\eta=0.0025$ is larger than the the IMF-weighted {\em stellar}
magnesium yield of $\sim 0.001$ \citep{saitoh}. Although numerical
simulations do require that a significant fraction of the stellar
yield is returned to the halo \citep{peeples_budget}, it is still the case
that a $10^8M_\odot$ stellar population produces too little \mgii to
fill a halo to 90kpc with observable \mgii, by a factor of a few.

This discrepancy is reduced if we invoke a larger number of lower-mass
halos, with radii scaled as $R\propto M^\gamma$ as calculated in
Figure 13. Then, the required yield becomes: 

\begin{equation} 
\eta = 2.5 \times 10^{-5} \left(\frac{M_*}{10^{10}
M_{\odot}}\right)^{2\gamma-1} \left(\frac{N_{\mmgii}}{10^{13} \text{cm}^{-2}}\right) \left(\frac{\kappa}{0.5}\right)
\left(\frac{f}{0.1}\right)^{-1}.  
\end{equation}

For $\gamma\sim0.39$ \citep{churchill_rvir} and a $M_*\sim 10^7M_\odot$
stellar mass (associated with the $M_h\sim 10^8M_\odot$), $\eta =
0.00011$, roughly an order of magnitude smaller than the IMF-weighted
magnesium yield, leaving a comfortable margin to account for the
difference between the strict stellar yield and the galactic yield of
Mg mass ejected into the halo.

As a final, crude consistency test, we explore whether individual small halos
have the correct combination of size and density to produce observable
\mgii absorption lines. For an average chord length throguh the halo
of $\bar{l}=\frac{4}{3}R$, and the mass-scaled halo radius relation
from above, one obtains an order-of-magnitude estimate of the column
density:
\begin{equation}
N_{\mmgii} =  4\times {10^{13} \text{ cm}^{-2}} \left(\frac{\eta}{0.001}\right) \left(\frac{f}{0.1}\right)^{-1}  \left(\frac{M_*}{10^{7} M_{\odot}}\right)^{1-2\gamma}.
\end{equation}

This column density is sufficient to produce saturated absorption,
though in any realistic model of the halo one expects cool gas to be
more highly structured\citep{crighton}, leading to a lower covering
fraction but slightly more variant total column densities in individual
sightline samples.

Taken together, these results point to a modest tension for models
where \mgii is hosted at high redshift by massive galaxies with $R\sim
100$ kpc gas envelopes; in contrast models populating \mgii in
galaxies with $\sim 10-100\times$ smaller stellar mass but
$3-10\times$ smaller gas envelopes comfortably accommodate the
observations for reasonable heavy element yields.

Such objects would be qualitatively distinct from the $L^*$ galaxies
hosting \mgii in the low-redshift universe, although they could evolve
over time into such massive systems.  If they are not yet dynamically
relaxed, it may be the case that the observed \mgii is not solely a
byproduct of winds from the halo's internal stellar population, but
rather combines winds with material stripped through interactions
during the initial assembly of the halo.  In this case some fraction
of the heavy elements producing observed absorption may never have
been in the halo center, reducing the requirements on wind transport
during epochs where the Hubble time was $<1$ Gyr.

\subsection{Limitations of Large-Scale Simulations for Interpreting \mgii Observables}

The statistics presented in the previous section made reference to
cosmological simulations of galaxy formation (specifically the
Illustris simulation) but employed a simple analytic model to predict
the likelihood of absorption by a given galaxy's CGM.  This model
utilizes covering fractions derived from low redshift observations to
derive a binomial hit/miss rate, and has no power to predict
equivalent widths or absorber kinematics (which are closely
correlated).

These same simulations incorporate sophisticated hydrodynamic solvers
and can therefore can be used---at least in principle---to calculate
line densities and frequency distributions directly without resort to
assumptions about covering fraction.  Indeed, these CGM statistics can
serve as an independent check on the simulations' feedback
prescriptions, beyond the present day galaxy mass function and
star-formation main sequence (which the simulations reproduce by
design). In practice however, computational limitations of the
simulations make direct predictions of the cosmological evolution of
cool gas quite difficult.  In this section, we use a simple analysis
of the Illustris simulation \citep{illustris, genel2015, sijacki2015}
to demonstrate some of the challenges.

\begin{figure}[h]
	\begin{minipage}{0.5\textwidth}
	\includegraphics[width=\textwidth]{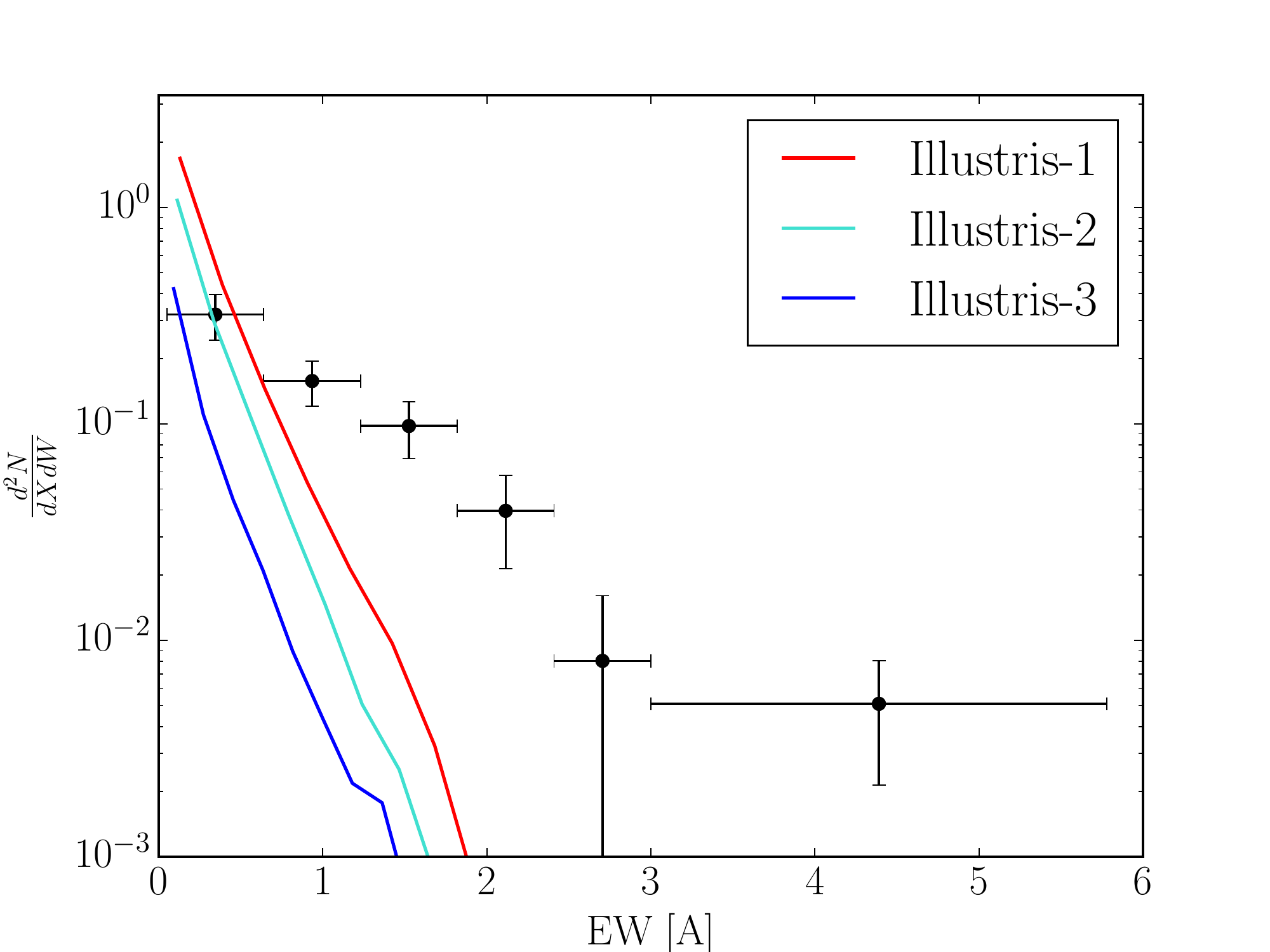}
	\caption{Comparison of the survey equivalent width distribution (blue points) to three different runs of Illustris at redshift z = 3.5 to test resolution convergence. The simulation boxes for Illustris 1, 2, and 3 are $75h^{-1}$ comoving Mpc on a side, with $18.1,2.3$ and $0.3\times 10^9$ hydro cells and a minimum cell size of $48, 98,$ and $273$ pc, respectively. Weaker absorbers with $W_r<1$\AA ~are produced at approximately the correct rate, although the equivalent width distribution is not fully converged even at the highest resolution.  Stronger absorbers are underproduced by a large factor at all resolutions.}
	\label{fig:fig_illustris}
	\end{minipage}
\end{figure}

Figure \ref{fig:fig_illustris} depicts the \mgii absorber frequency
distribuion in our $z=3.15$-$3.81$ redshift bin, along with the
predicted frequency distributions found in three different runs at
these redshifts used for resolution convergence testing in Illustris.
The simulation boxes are $75h^{-1}$ comoving Mpc on a side, with
$18.1,2.3$ and $0.3\times 10^9$ hydro cells and a minimum cell size of
$48, 98,$ and $273$ pc, respectively \citep{illustris}.  The
simulation tracks metallicity, temperature, density, and velocity for
each cell in the simulation. We calculate absorption profiles in
post-processing using the methodology and code described in
\citet{bird_spectra}.  In short, the ionization balance for each cell
is calculated using the UV background spectrum of \citet{faucher_uvbg}
at the appropriate redshift, applied to a grid of ionization fractions
calculated using CLOUDY \citep{cloudy}. Because \mgii has an
ionization potential of 1.1 Ryd, neutral hydrogen can shield it
efficiently from ionizing radiation.  Since its ionization potential
is very close to that of hydrogen, we make a simple correction for
self-shielding of absorbing structures using the formalism of
\citet{rahmati}. Absorbers were identified via instances where the simulated spectra
dip 5 percent below continuum values. Absorption troughs within 500 km/s of
each other were grouped together and identified as single absorbers.
The equivalent widths of such absorbers were then calculated by integrating
500 km/s past the most extremal components of each absorber. 

Figure \ref{fig:fig_illustris} shows that the simulations produce too
few \mgii absorbers except for the weakest values of $W_r$.  There is
a marginal increase in the normalization of the predicted frequency
distribution as the simulation resolution is increased.  However, the
slope of the simulated distribution remains steeper than the observed
slope at all resolutions.  Even at $W_r\sim1$\AA, the Universe
has $3$-$10\times$ more absorbers than the simulations; at larger
equivalent widths the discrepancy spans many orders of magnitude,
since the box contains few or no systems at $W_r\gtrsim 2$\AA.

The simulations' relatively coarse mesh resolution, required to
simulate a large cosmological volume, likely contributes to this
deficit of strong absorbers. 
The spatial resolution of a single particle can be approximated as
\begin{equation}
l = 7.1 \mathrm{kpc} \left( \frac{m_{\mathrm{res}}}{10^6 M_\odot} \right)^{1/3}  \left( \frac{\rho}{10^{-3}\mathrm{cm}^{-3}} \right)^{-1/3}
\end{equation}
where $m_{\mathrm{res}}$ is the gas mass resolution of the simulation and $\rho$ is the gas density.
Even with the aggressively high particle count in the Illustris simulation, the baryon mass resolution is $m_{\mathrm{res}}=1.26 \times 10^6 M_\odot$ with gives an associated particle spatial resolution of $\sim$7 kpc.
While the dense interstellar medium will have higher spatial resolution, the comparatively diffuse CGM can only be expected to have spatial resolution of order tens of kiloparsecs.
The coarse resolution can affect
predictions of absorption statistics in two important ways.

First, limited simulation spatial resolution may suppress important
small scale fluctuations in density, temperature, and velocity.  In
the case of the CGM in Illustris-1, any density/temperature
fluctuations or bulk/turbulent velocity fluctuations with spatial
scales less than $\sim$10 kpc at typical CGM gas densities will not be
captured.  For the lower resolution Illustris-2 and Illustris-3
simulations, the CGM spatial resolution is degraded further.  Indeed
the higher resolution runs exhibit an increase in the overall
normalization of the \mgii frequency distribution, indicating that
numerical convergence has not yet been attained in regions producing
\mgii absorption. Moving to yet higher resolution may lead to a
continued increase in the overall normalization of the \mgii frequency
distribution as smaller-scale density fluctuations continue to be
resolved
\footnote{We note, however, that in addition to resolving
  smaller-scale density fluctuations, moving to higher resolution also
  changes the total amount of Mg in the simulations volume, as the
  stellar mass (and therefore the total metal mass) is not fully
  converged even at Illustris-1 resolution.}.

Evidence for unresolved simulated CGM velocity substructure comes 
from the deficit of strong absorbers.  In these systems, individual absorption components are saturated, so the \mgii equivalent width is a proxy for velocity dispersion.  Simple tests show that the Illustris \mgii absorbers easily meet the column density criterion for saturation.  They simply did not have enough velocity substructure to generate large equivalent widths. The existence of unresolved simulated CGM velocity substructure could easily follow from the limited spatial resolution employed in Illustris. As with density fluctuations, bulk/turbulent velocity distributions on spatial scales less than a few cell sizes (i.e. tens of kpc) are unlikely to be captured.

\begin{figure}[h]
	\begin{minipage}{0.5\textwidth}
	\includegraphics[width=\textwidth]{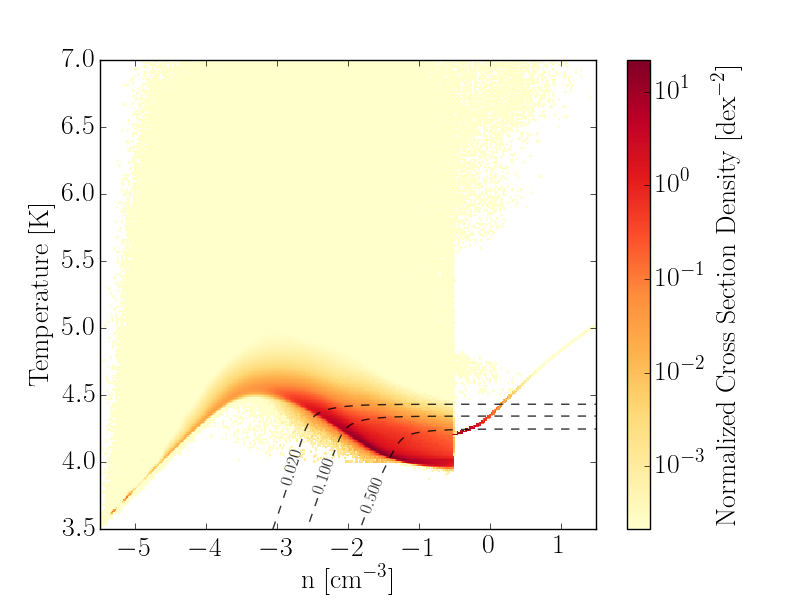}
	\caption{Temperature-density phase diagram of illustris cells,
          color coded by cross-section weighted \mgii distribution function.  Dashed lines show the \mgii ionization fraction $f_{\mgii}$. The $f_{\mgii} = 0.1$ contour, roughly corresponding to temperatures below $10^{4.5}$ K and densities above $10^{-2} \text{ cm}^{-3}$, contain about 82\% of the \mgii in Illustris. Most of the \mgii thus reside in either under-resolved regions of the simulation or on the effective equation-of-state.}
	\label{fig:phase_diag}
	\end{minipage}
\end{figure}

The second and more subtle issue concerns the position of \mgii
absorbing gas in the density-temperature plane.  As a ubiquitous
constituent of the CGM, \mgii traces material in the transition range
between the IGM and the ISM.  To simplify and speed calculations in
these regions with short cooling time scales, most numerical codes
also transition somewhere in this density regime to a sub-grid physics
model that will necessarily compromise some of their predictive power
for studying gas physics.  Figure \ref{fig:phase_diag} shows the phase
diagram in the Temperature-density plane for Illustris, with a color
scale indicating the cross-section weighted \mgii distribution function in different parts of the plane (i.e. the fraction of all \mgii cross section in the box contained within a bounded logarithmic interval of $n$ and $T$). More than 80\% of the aggregate \mgii cross section in Illustris resides within the contour defined by an ionization-fraction $f_{\mmgii} = 0.1$. Here gas is either generally under-resolved (Equation 13) or has transitioned onto an effective equation-of-state implemented in Illustris (represented by a thin line at lower right for $n>0.13$ cm$^{-3}$).

Particles on the effective equation-of-state are modeled using a two phase
medium of cool clouds embedded in a hot tenuous phase, and these
particles are assigned a star formation rate and associated IMF.  In
practice, these cells are treated as current or future ISM
constituents, and simplified in exactly the region where the phase
diagram indicates that \mgii should be strongest in the CGM.

These shortcomings could possibly be mitigated for the study of
individual absorbers by using a galaxy scale simulation with similar
resolution and/or refinement
\citep{eris,churchill_simulations,FIRE_feedback}.  By transitioning to
sub-grid physics at smaller scales these smaller boxes track CGM gas
physics with higher fidelity but are less useful for studying
cosmological statistics and number counts. 

\begin{figure}[h]
	\begin{minipage}{0.5\textwidth}
	\includegraphics[width=\textwidth]{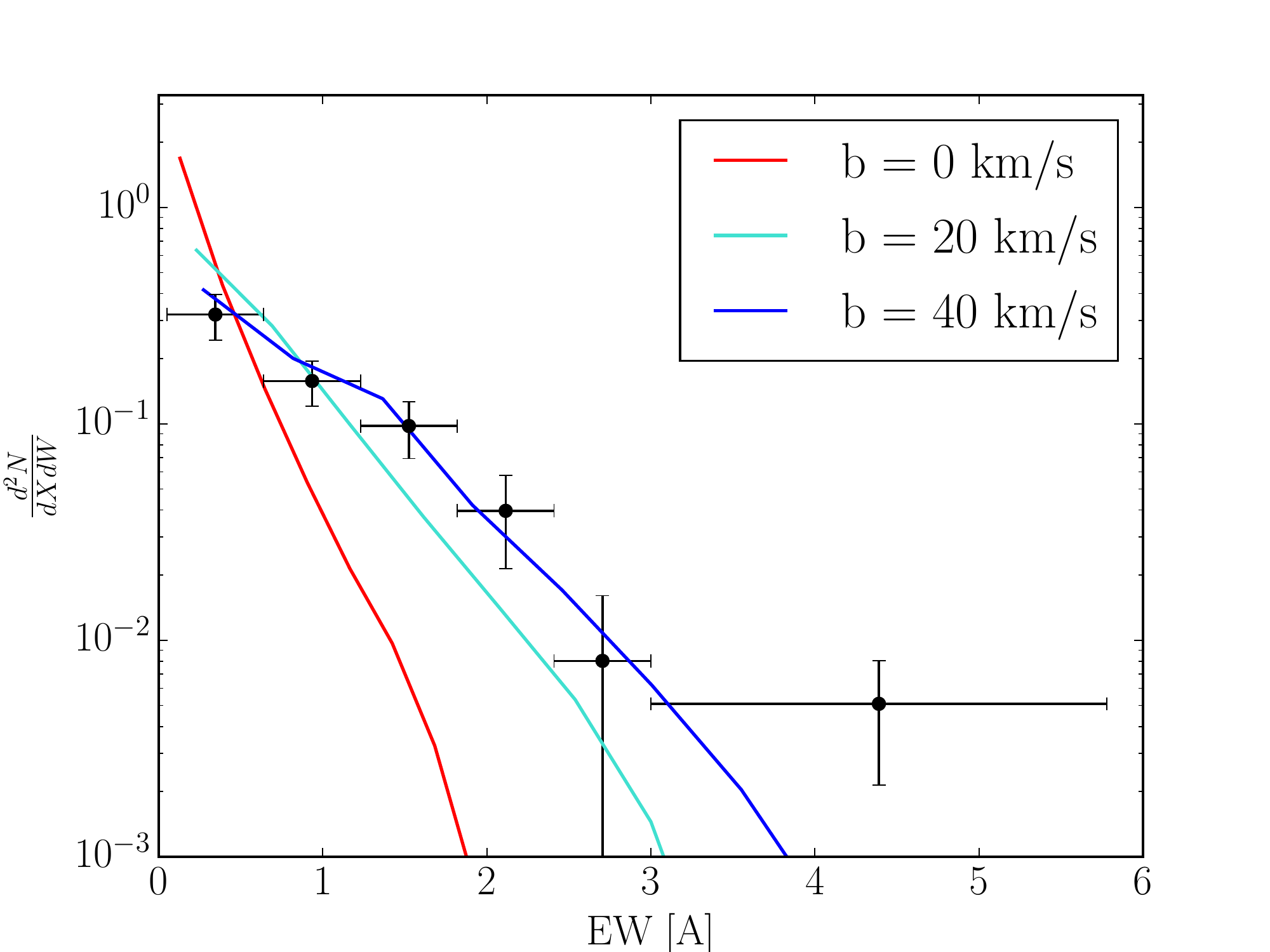}
	\caption{Reconciliation of the equivalent width distribution
          determined from Illustris sightlines (solid lines) with
          observed data at $z=3.5$ (points with errors) by 
          artificial inflation of the 1D bulk velocity dispersion.  
          The paucity of strong absorbers in
          the un-adjusted simulation arises from unresolved velocity
          substructure rather than insufficient densities---the
          simulated densities produce saturated absorption, just over
          too small of a velocity interval.  Artificial inflation with
          a $40$ km/s kernel brings the two curves into agreement.}
	\label{fig:turb}
	\end{minipage}
\end{figure}

As an intermediate step, we tested whether the observed equivalent
width distribution could be reproduced by artificially inflating the
line-of-sight velocity dispersions of Illustris \mgii in
post-processing.  We find that this can be achieved by uniformly
broadening the Voigt profiles of simulation particles contributing to
our simulated spectra, effectively adding an additional velocity dispersion associated with unresolved bulk flows of $b \sim 40 $ \kms (Figure~\ref{fig:turb}).  This
procedure has no effect on the equivalent width of unsaturated
profiles, but for saturated systems $W_r$ is increased by spreading
the absorption over a wider velocity range.  

The driving source for this unresolved turbulence is left unspecified.
However the strongest ($W_r>1$\AA) absorbers that most clearly reveal
unresolved velocity substructure peak in incidence at $z=2$-$3$,
coincident with the global star formation rate (Figure \ref{fig:sfr}).  

\begin{figure}[h]
	\begin{minipage}{0.5\textwidth}
	\includegraphics[width=\textwidth]{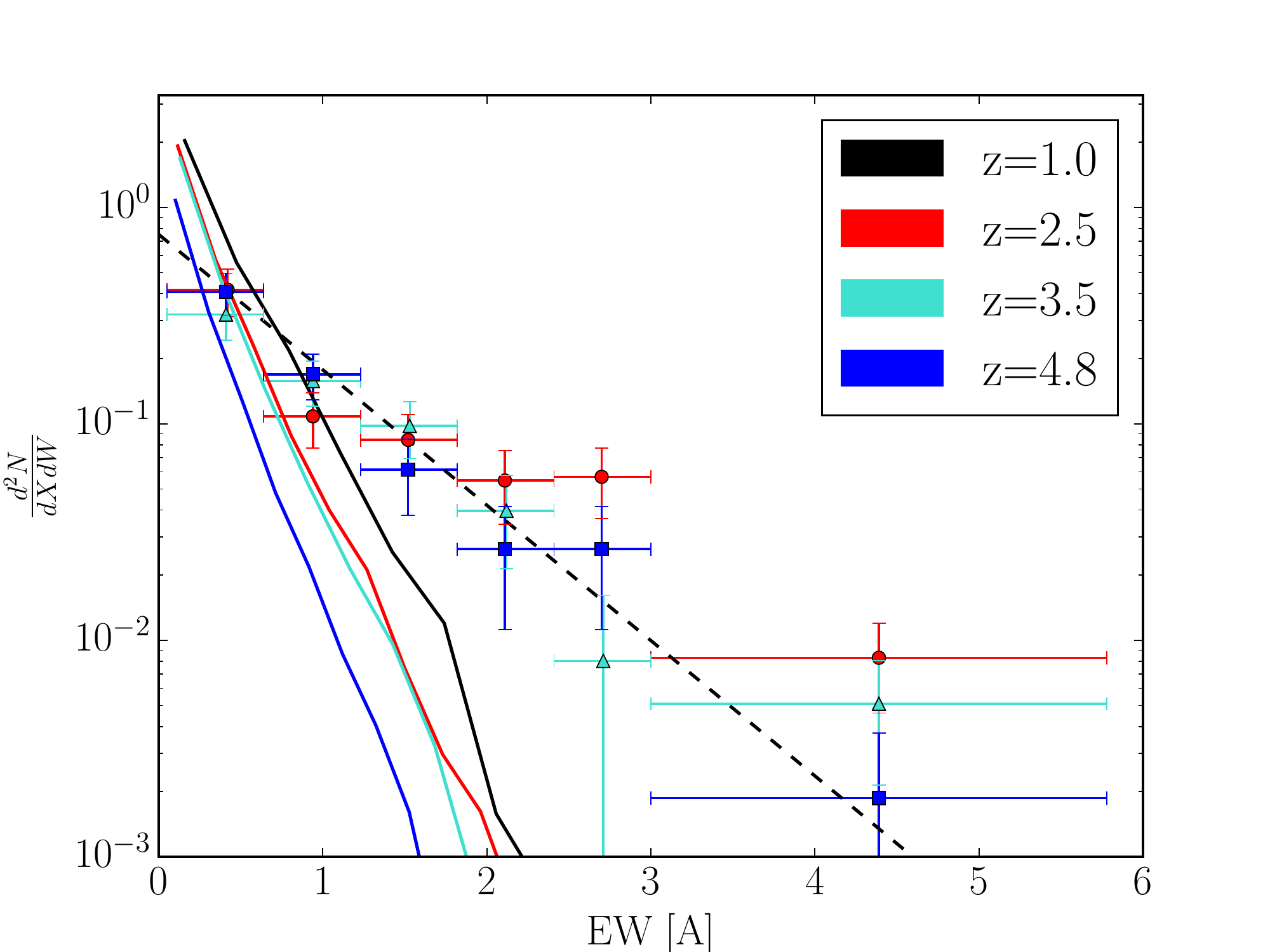}
	\caption{Redshift evolution of the equivalent width
          distribution both from observations (colored points) and
          Illustris simulations (solid curves).  At all redshifts, the
          simulated slope is too steep (see Figure \ref{fig:turb}).
          Also, there is mild evolution in the normalization of the
          simulated curves, in contrast to the data which evolve
          somewhat in slope but not in normalization.}
	\label{fig:z_evol}
	\end{minipage}
\end{figure}

Although this prescription works for any single redshift, Illustris
also predicts redshift evolution in the overall normalization of the
frequency distribution, in contrast to the observations (Figure \ref{fig:z_evol}).  Yet despite
these detailed discrepancies, the simulations do produce approximately
the correct total number of \mgii absorbers---they simply allocate too
many to low values of $W_r$.  A natural interpretation is that Mg and
even \mgii is broadly being produced and distributed in a physically
plausible spatial pattern within the simulation box, but is
insufficiently stirred.  The large-volume simulations would then
provide accurate overall number counts, but smaller volumes
\citep{joung,survival_clouds,bruggen} will be the most promising
avenue to resolve outstanding questions about the micro-physics of
enrichment, turbulence, and the interaction of intra-halo gas with
material flowing into and out of the central disk.

\section{conclusions}

We have completed an infrared survey for \mgii at $1.9<z<7.0$,
augmenting the sample first presented in Paper I.  We searched
Magellan/FIRE spectra of \NumQSOs ~QSOs, with particular emphasis in
this paper on objects at $z>6$, which provide \mgii pathlength in the
$K$ band that was not explored in previous work.  This choice enables
a more significant detection of evolutionary trends in the \mgii
absorber population. The reported \mgii absorbers were identified by
means of an automated finding routine and verified by eye; the
completeness and false-positive rates of both automated and
user-evaluated steps were then tested using a large suite of Monte
Carlo simulations. Our main findings can be summarized as follows:

\begin{enumerate}

\item
We confirm and strengthen an evolutionary decline in the frequency of
strong \mgii absorbers ($W_r > 1$ \AA) by roughly an order of
magnitude in our survey's redshift range, while the frequency of
weaker absorbers remains remarkably constant from $z=0.25$-$7.0$. The
\mgii equivalent width distribution function (EWDF) slope $W_{\ast}$
does not continue its growth at low redshifts but rather peaks at $z
\sim 2$-$3$, after which the incidence of strong absorbers begins to
decline.

\item
The inclusion of high-redshift sightlines yielded \NumHighZ ~systems
with $z > 6$. These are the first known \mgii absorbers above this
redshift, which was not covered in Paper I.  Remarkably, these
detections are consistent with the continued non-evolution of the weak
absorbers seen at low redshifts. The single strong absorber detected
at $z > 6$ is also consistent with the decline of the strong absorber
population, albeit with large shot noise. The persistence of the low-z
population trends is also reflected in the continued decay of
$W_{\ast}$ from $z = 2$ onwards.

\item
The peak in the frequency of strong \mgii absorbers at $z \sim 2$ and
the subsequent decline towards higher redshifts resembles
the evolution of the cosmic SFR density. Our new measurements confirm
the agreement between the \mgiinsp-to-SFR density scaling of
\citep{menardo2} and direct measurements of the SFR density using the
Hubble Space Telescope. When this analysis is extended to include our
highest redshift strong absorber, we find that the \mgiinsp-derived
and observed SFR densities are coincident. Aforementioned cautions
aside, the agreement between thus derived SFR densities suggests a
connection between \mgii absorbers and star formation.

\item
Analytic calculations using Illustris-derived halo mass functions,
which populate halos above a lower mass bound prescripteively with
\mgiinsp, severely under-predict the total incidence of absorbers with
$W_r > 0.3$ \AA ~at $z\gtrsim 2.5$. This is unsurprising given the
non-evolution of the weak absorbers and the substantial evolution of
the mass function in our survey's redshift range. A simple
modification wherein we populate halos above a minimum SFR cut (rather
than a mass cut) achieves a redshift-dependent minimum mass.  This
partly alleviates the models' deficiency of \mgii at high redshift but
remains rather imperfect.

\item
Spectra computed directly by projecting sightlines through the
Illustris simulation's gas volume produce approximately the correct
number of weak \mgii absorbers but they under-predict strong \mgii
counts at all redshifts. Current cosmological simulations do not
sufficiently resolve the structure of CGM gas to properly sample the density and
temperature regime at which \mgii absorption manifests. As both
simulated and observed \mgii absorbers tend to be saturated, subgrid
models which merely add spatial power or increase the simulated
\mgii column density cannot reproduce the observed \mgii absorber
distribution.  Because equivalent widths reflect turbulent broadening
of saturated components, a successful subgrid model will need to
include velocity substructure as well. We are able to qualitatively
reproduce the observed \mgii rest equivalent width distribution by
assuming subgrid bulk turbulence at the $b \sim 40$ \kms level in
Illustris; however the degree of artificial turbulence required is
specific to the simulation being considered.

\end{enumerate}

The continued presence of \mgii absorption at $z=6$-$7$---even in our
limited number of $z>6.3$ sightlines---suggests that the enrichment of
the CGM was underway quite early, while the dark matter halos of
future galaxies were still assembling and their stellar populations
had yet to fully form and coalesce into pronounced disks.  Indeed our
highest redshift point (centered at $z=6.3$) post-dates the
instantaneous reionization redshift inferred from the $\tau=0.058$
measurement of Planck \citep{planck_tau} by only $240$ Myr.  Given
that matter moving at 100 \kms travels roughly 1 kpc every 10 Myr,
there is scarcely enough time for a halo to accrete gas into its
center, develop a stellar population, deposit feedback into the
surroundings, and transport such material far enough back into the
halo to produce an appreciable absorption cross section. It may be the
case that early intra-halo gas is enriched by elements manufactured in
accreted satellites or other in-situ star formation environments
before they are subsumed into a galaxy's central condensate.  This
would especially be true if early \mgii absorbers preferentially occur
in rich or highly biased environments where the earliest galaxies
would have formed.

Our detection of numerous \mgii systems at $z>6$---unlike \civ systems
which decline rapidly toward higher $z$ \citep{simcoeCiv}---is
promising for future investigation of the reionization epoch using
metal absorption lines.  Searches for low-ionization metals in \oinsp,
\siiinsp, and \cii in the $J$ band have yielded some success
\citep{becker_oi}, but these ions cover a smaller pathlength per QSO
because they have rest wavelengths near \lyansp.  \mgii is an abundant
$\alpha$ element with large oscillator strength, has a long
pathlength, is easily identified as a doublet, and can be measured out
to $z\sim 8$ from the ground if suitable QSOs are identified.  If so,
it will be possible to study metal enrichment in CGM from nearly the
current epoch ($z=0.2$) to the CMB electron scattering redshift
($z\sim 8$) using a single characteristic transition.

\acknowledgements

It is a pleasure to thank the staff of the Magellan Telescopes and Las
Campanas Observatory for their hospitality and assistance in obtaining
the data described here. 
We thank Hsiao-Wen Chen for useful discussions.
The computations in this paper were run in part on the Odyssey cluster supported by the FAS Division of Science, Research Computing Group at Harvard University. This material is based upon work supported by the National Science Foundation Graduate Research Fellowship under Grant No. DGE 1106400.
We gratefully acknowledge direct funding
support from the MIT Undergraduate Research Opportunity (UROP) program
(SC) and NSF award AST-0908920.
PT acknowledges support from the Hubble Fellowship (HST-HF2-51384.001-A).  Support for program number HST-HF2-51384.001-A was provided by NASA through a grant from the Space Telescope Science Institute, which is operated by the Association of Universities for Research in Astronomy, Incorporated, under NASA contract NAS5-26555.
KLC acknowledges support from AST-1003139, which funded her participation
in this work. RS thanks the Radcliffe Institute of Advanced Study for
their support and hospitality during the final phase where this paper
was completed.
We collectively recognize support from the Adam
J. Burgasser Chair in Astrophysics.

\bigskip

\facility{Magellan:FIRE}

\bibliographystyle{aasjournal}
\bibliography{mgii}{}

\clearpage

\end{document}